\documentclass[titlepage]{article}
\usepackage{amssymb}

\usepackage{amsmath}
\usepackage{hyperref}
\usepackage{afterpage}

\begin{document}

\title{ Dynamical Charged N-body Equilibrium in Circular Dilaton Gravity}
\author{Ryan Kerner and Robert B. Mann}
\maketitle

{\LARGE Abstract}

We extend the problem of $(1+1)$ circular dilaton gravity to include charged
particles.  We examine the two (charged) particle case in detail and find
an exact equilibrium solution. We then extend this to $N-$particles and
obtain a solution for this case as well. This class of solutions corresponds
to $N$-particles of the same mass, spaced evenly around the circle with
charges chosen so that the electric field satisfies $E^{2}=$constant.  We
discuss the relation of these solutions to the previous uncharged
equilibrium solutions and examine the behavior when the number of particles
is large. We comment on the challenges in further generalizing the solutions
we obtain.

\section{Introduction}

Circular gravity is a general relativistic gravitational system in $(1+1)$
dimensions in which the topology of space is equivalent to that of a circle $%
S^{1}$. \ Its counterpart, lineal gravity (the same system with the $S^{1}$
topology replaced by $\mathbb{R}^{1}$), has been the subject of much
investigation over the past twenty years, motivated by the desire to
understand black holes \cite{r3}, space-time structure \cite{browndan,Strobl}%
, canonical quantum gravity \cite{Gabor}, string physics \cite{Witten},
information loss in black hole evaporation \cite{CGHS} and the $N-$body
problem \cite{OR,pchak}.

This latter problem involves consideration of a collection of $\ N$
particles interacting through their own mutual gravitational attraction,
along with other specified forces. \ Such systems are referred to as
self-gravitating systems. Apart from functioning as prototypes for higher
dimensional gravity, in lineal gravity they also approximate the behavior of
some physical systems in three spatial dimensions, such as stellar dynamics
orthogonal to a galactic plane, the collisions of flat parallel domain
walls, and the dynamics of cosmic strings. The ergodic and equipartition
properties of non-relativistic one-dimensional self-gravitating systems are
a current subject of research \cite{yawn}, and recently such systems were
shown to exhibit fractal behavior \cite{fractal}. \ 

In the past few years these systems have been extended to relativistic
lineal gravity. The 2-body problem has been solved exactly \cite{2bd}, as
have its extensions to include cosmological expansion \cite%
{2bdcossh,2bdcoslo}\ and/or electromagnetic interactions \cite{2bdchglo}. \
\ These extensions, while considerably more complicated, are nevertheless
amenable to statistical-mechanical analysis \cite{pchak}. They include the
non-relativistic self-gravitating systems \cite{yawn,Rybicki} as a limiting
case. \ 

Comparatively little is known about the circular counterparts of these
lineal systems. Preliminary investigations \cite{circle} have shown that
qualitatively new features emerge that are absent in the non-compact lineal
case. \ For comparison, in the absence of other fields there are no static
non-relativistic equilibrium solutions for $N$ particles on a circle. This
is because the non-relativistic potential grows linearly with increasing
distance from the source, and a non-relativistic circular topology admits no
solutions to the matching conditions for any $N\geq 1$. \ However it has
recently been shown that the relativistic case has an exact equilibrium
solution for $N$ particles \cite{rkern}. This solution is possible because
the space-time itself either expands or contracts in response to the
sources, an option not available in the non-relativistic case.

In this paper we extend the formalism of the $N$-body problem in circular
gravity to include electromagnetic interactions. We search for equilibrium
solutions in the case when the particles are charged. \ We examine the
charged 2-body system in detail, and find a class of charged solutions that
can be generalized to $N$-particles and encompasses the previous non-charged
solutions. Our solutions are not the most general, and we comment on the
challenges involved in extending the solutions that we obtain.

\bigskip

\section{Canonical Circular Gravity}

We begin with the general framework previously formulated for canonical
circular gravity minimally coupled to $N$ particles in the presence of a
cosmological constant $\Lambda $ \cite{circle}, extending the formalism to
include coupling to electromagnetism. \ This can be done in a manner
analogous to that way that charges were added to lineal gravity \cite%
{2bdchglo}. \ The resulting action is \ 

\begin{eqnarray}
I &=&\oint d^{2}x\left[ \frac{1}{2\kappa }\sqrt{-g}g^{\mu \nu }\left\{ \Psi
R_{\mu \nu }+\frac{1}{2}\nabla _{\mu }\Psi \nabla _{\nu }\Psi +\frac{1}{2}%
g_{\mu \nu }\Lambda \right\} \right.  \notag \\
&&-A_{\mu }F_{,\nu }^{\mu \nu }+\frac{1}{4\sqrt{-g}}F^{\mu \nu }F^{\alpha
\beta }g_{\mu \alpha }g_{\nu \beta }  \notag \\
&\qquad &+\sum_{a}\int d\tau _{a}\left\{ -m_{a}\left( -g_{\mu \nu }(x)\frac{%
dz_{a}^{\mu }}{d\tau _{a}}\frac{dz_{a}^{\nu }}{d\tau _{a}}\right)
^{1/2}+e_{a}\frac{dz_{a}^{\mu }}{d\tau _{a}}A_{\mu }(x)\right\}  \notag \\
&&\left. \times \delta ^{2}(x-z_{a}(\tau _{a}))\right] ,  \label{act0}
\end{eqnarray}%
where $\Psi $ is the dilaton field, $A_{\mu }$ and $F^{\mu \nu }$ are the
vector potential and the field strength, $g_{\mu \nu }$ and $g$ are the
metric and its determinant, $R$ is the Ricci scalar, and $\tau _{a}$,$z_{a}$%
, and $e_{a}$ are the proper time, position, and charge of the $a$-th
particle, with $\kappa =8\pi G/c^{4}$. The range of the spatial coordinate $%
x $ is $-L\leq x\leq L$, and the circular topology implies that all fields
must be at least $C^{1}$ functions of $x$ with period $2L$, therefore

\begin{equation}
f(L)=f(-L)\qquad \text{and}\qquad f^{\prime }(L)=f^{\prime }(-L)\;\;.
\label{smooth}
\end{equation}

The action (\ref{act0}) describes a generally covariant self-gravitating
system (without collisional terms, so that the bodies pass through each
other); its field equations are

\begin{align}
& R-g^{\mu \nu }\nabla _{\mu }\nabla _{\nu }\Psi =0\ ,  \label{eq-R} \\
& \frac{1}{2}\nabla _{\mu }\Psi \nabla _{\nu }\Psi -\frac{1}{4}g_{\mu \nu
}\nabla ^{\lambda }\Psi \nabla _{\lambda }\Psi +g_{\mu \nu }\nabla ^{\lambda
}\nabla _{\lambda }\Psi -\nabla _{\mu }\nabla _{\nu }\Psi =\kappa T_{\mu \nu
}+\frac{1}{2}g_{\mu \nu }\Lambda \;,  \label{eq-Psi} \\
& F_{,\nu }^{\mu \nu }=\sum_{a}e_{a}\int d\tau _{a}\frac{dz_{a}^{\mu }}{%
d\tau _{a}}\delta ^{2}(x-z_{a}(\tau _{a})),  \label{eq-F} \\
& \frac{1}{\sqrt{-g}}F_{\mu \nu }=\partial _{\mu }A_{\nu }-\partial _{\nu
}A_{\mu }, \\
& m_{a}\left[ \frac{d}{d\tau _{a}}\left\{ g_{\mu \nu }(z_{a})\frac{%
dz_{a}^{\nu }}{d\tau _{a}}\right\} -\frac{1}{2}g_{\nu \lambda ,\mu }(z_{a})%
\frac{dz_{a}^{\nu }}{d\tau _{a}}\frac{dz_{a}^{\lambda }}{d\tau _{a}}\right] 
\notag \\
& =e_{a}\frac{dz_{a}^{\nu }}{d\tau _{a}}\{A_{\nu ,\mu }(z_{a})-A_{\mu ,\nu
}(z_{a})\}\;,  \label{eq-z3}
\end{align}%
where the stress-energy due to the point masses and the electric field is%
\begin{align}
T_{\mu \nu }& =\sum_{a}m_{a}\int d\tau _{a}\frac{1}{\sqrt{-g}}g_{\mu \sigma
}g_{\nu \rho }\frac{dz_{a}^{\sigma }}{d\tau _{a}}\frac{dz_{a}^{\rho }}{d\tau
_{a}}\delta ^{2}(x-z_{a}(\tau _{a}))  \label{stress} \\
& +\frac{1}{-g}\{F_{\mu \alpha }F_{\nu \beta }g^{\alpha \beta }-\frac{1}{4}%
g_{\mu \nu }F_{\alpha \beta }F^{\alpha \beta }\},  \notag
\end{align}

Recall that there is no magnetic component of the field in $(1+1)$
dimensions. Conservation of $T_{\mu \nu }$ is ensured by eq. (\ref{eq-Psi}).
\ Inserting the trace of eq. (\ref{eq-Psi}) into eq. (\ref{eq-R}) yields

\begin{equation}
R-\Lambda =\kappa T_{\;\;\mu }^{\mu }\;.  \label{RT}
\end{equation}

Consequently eqs. (\ref{eq-F})-(\ref{eq-z3}) and (\ref{RT}) form a closed
system of equations for gravity, matter, and electromagnetism leaving\ (\ref%
{eq-Psi}) to determine the evolution of the dilaton field $\Psi $. \ \ The
preceding system generalizes Jackiw-Teitelboim lineal gravity \cite{JT},
which equates scalar curvature to a (cosmological) constant $\Lambda $,
reducing to it if the stress-energy (\ref{stress}) vanishes. The specific
choice of dilaton coupling in $R=T$ \ theory is such that the evolution of
the dilaton does not modify the aforementioned reciprocal
gravity/matter/electromagnetic dynamics. \ The total system is a closed
system that describes the evolution of the single metric degree of freedom
and the $N$ degrees of freedom of the charged point masses. The evolution of
the charged point-masses governs the evolution of the dilaton via (\ref%
{eq-Psi}), whose divergencelessness is consistent with the conservation of $%
T_{\mu \nu }^{P}$, yielding only one independent equation to determine the
single degree of freedom of the dilaton.

The metric is written as $ds^{2}=-\left( N_{0}dt\right) ^{2}+\gamma \left(
dx+\frac{N_{1}}{\gamma }dt\right) ^{2}$, where $N_{0}$ is the lapse and $%
N_{1}$ is the shift. \ With this metric the extrinsic curvature $K$ is then 
\begin{equation}
K=(2N_{0}\gamma )^{-1}(2\partial _{1}N_{1}-\gamma ^{-1}N_{1}\partial
_{1}\gamma -\partial _{0}\gamma )=\sqrt{\gamma }\kappa (\pi -\Pi /\gamma )
\label{kext}
\end{equation}%
where the conjugate momenta to $\gamma $ and $\Psi $ are respectively
denoted $\pi $ and $\Pi $. \ We can take the extrinsic curvature $K$ to be a
time coordinate $\tau \left( t\right) $ of the circle, thereby allowing the
elimination of $\pi $ from all field equations since $\pi =\frac{\tau }{%
\kappa \sqrt{\gamma }}+\frac{\Pi }{\gamma }$.

Simplifying the action (\ref{act0}) involves decomposing the scalar
curvature in terms of the extrinsic curvature via the equation $\sqrt{-g}%
R=-2\partial _{0}(\sqrt{\gamma }K)+2\partial _{1}[(N_{1}K-\partial
_{1}N_{0})/\sqrt{\gamma }$. \ Substituting this result into the action (\ref%
{act0}), and eliminating \ $\pi $ via $\pi =\frac{\tau }{\kappa \sqrt{\gamma 
}}+\frac{\Pi }{\gamma }$ yields \ \cite{2bdchglo} \cite{circle}%
\begin{equation}
I=\oint d^{2}x\left\{ \sum_{a}p_{a}\dot{z}_{a}\delta (x-z_{a})+\Pi \frac{%
\partial }{\partial t}\left( \Psi +\ln \gamma \right) -\mathcal{H}\right\}
\;,  \label{act3}
\end{equation}%
where $H=\oint dx\mathcal{H=}\frac{2\dot{\tau}}{\kappa }\oint dx\sqrt{\gamma 
}$. \ $H$ is the circumference functional of the circle when $\dot{\tau}$ is
constant. The dynamics is that of a time-dependent system, with the
time-dependence corresponding to the time-varying circumference of the
circle of constant mean extrinsic curvature. As with general relativity on
spatially compact manifolds in $(2+1)$\ dimensions \cite{Moncreif}, \ the
spacetime metric induces the metric $d\sigma ^{2}=$\ $g_{11}\left(
x,t\right) dx^{2}$on each $t=$constant hypersurface. Since any non-singular
metric on a circle is globally conformal to a flat metric, we choose the\
spatial metric to be such that $\gamma =\gamma (t)$\ , i.e. \ $\gamma
^{\prime }=0$. \ In other words, we write the $g_{11}$\ component of the
metric as $\gamma (t)h(x,t)$, and then set $h(x,t)=1$\ by a coordinate
transformation. By choosing a time parametrization so that $\dot{\tau}\sqrt{%
\gamma }$\ is constant, the Hamiltonian can be made time-independent.

It is important to note that one of the intermediate steps in the previous
simplification is a constraint equation on the electric field $E$ \cite%
{2bdchglo}, which is 
\begin{equation}
\frac{dE}{dx}=\sum_{a}e_{a}\delta (x-z_{a}(t))  \label{E-constr}
\end{equation}%
The solution to this constraint equation (\ref{E-constr}) is%
\begin{equation}
E=\frac{1}{2}\sum_{a}e_{a}\epsilon (x-z_{a}(t))+E_{0}  \label{E-field}
\end{equation}%
where $\epsilon (x)$ is the step function $\epsilon (x)=\frac{|x|}{x}$, and $%
\epsilon (0)=0$, or more explicitly 
\begin{equation*}
\left\{ 
\begin{array}{c}
\epsilon (x)=-1,x<0 \\ 
\epsilon (x)=0,x=0 \\ 
\epsilon (x)=1,x>0%
\end{array}%
\right\}
\end{equation*}

We chose a regularization such that $f(z_{a})=1/2(f(z_{a}+\epsilon
)+f(z_{a}-\epsilon ))$\ for any quantity that has a jump discontinuity. This
is necessary as otherwise it would not be possible to satisfy field
equations that depend on the specific value of the fields at the location of
the particles (i.e. field equations (\ref{fin15}),(\ref{fin15a}) or
equivalently (\ref{final-7}), (\ref{final-8}).). \ The choice of $\epsilon
=0 $\ at the particle is a reflection of this regularization and is
necessary to form a solution. \ \ 

Applying the smoothness condition across the boundary (\ref{smooth}) to the
equation for the electric field (\ref{E-field}) implies that $%
\sum_{a}e_{a}=0 $. \ So the total charge of the $(1+1)$ circular gravity
system must equal zero. \ 

The resulting field equations from the action (\ref{act3}) therefore become%
\begin{align}
& 2\Pi ^{\prime }-\Pi \Psi ^{\prime }+\sum_{a}p_{a}\delta (x-z_{a}(x^{0}))=0
\label{fin-9} \\
& \Psi ^{\prime \prime }-\frac{1}{4}(\Psi ^{\prime })^{2}-(\kappa \Pi
)^{2}+\gamma \left( \tau ^{2}-\Lambda /2+\kappa E^{2}/2\right)  \notag \\
& +\kappa \sum_{a}\sqrt{p_{a}^{2}+\gamma m_{a}^{2}}\;\delta
(x-z_{a}(x^{0}))=0,  \label{fin10} \\
& N_{0}^{\prime \prime }-\dot{\tau}\gamma  \notag \\
& -N_{0}\left\{ \gamma (\tau ^{2}-\Lambda /2+\kappa E^{2}/2)+\kappa \sum_{a}%
\frac{\gamma \;m_{a}^{2}}{2\sqrt{p_{a}^{2}+\gamma m_{a}^{2}}}\;\delta
(x-z_{a}(x^{0}))\right\} =0,  \label{fin11} \\
& N_{1}^{\prime }-\dot{\gamma}/2+\gamma \tau N_{0}=0,  \label{fin12} \\
& \dot{\Pi}+\partial _{1}(-\frac{1}{\gamma }N_{1}\Pi +\frac{1}{2\kappa \sqrt{%
\gamma }}N_{0}\Psi ^{\prime }+\frac{1}{\kappa \sqrt{\gamma }}N_{0}^{\prime
})=0,  \label{fin13} \\
& \dot{\Psi}+2N_{0}\left( \kappa \frac{\Pi }{\sqrt{\gamma }}+\tau \right)
-N_{1}(\frac{1}{\gamma }\Psi ^{\prime })=0,  \label{fin14} \\
& \dot{p}_{a}+\frac{\partial N_{0}}{\partial z_{a}}\sqrt{\frac{p_{a}^{2}}{%
\gamma }+m_{a}^{2}}-\frac{N_{0}}{2\sqrt{\frac{p_{a}^{2}}{\gamma }+m_{a}^{2}}}%
\frac{p_{a}^{2}}{\gamma ^{2}}\frac{\partial \gamma }{\partial z_{a}}-\frac{%
\partial N_{1}}{\partial z_{a}}\frac{p_{a}}{\gamma }+N_{1}\frac{p_{a}}{%
\gamma ^{2}}\frac{\partial \gamma }{\partial z_{a}}  \notag \\
& +\int dxN_{0}\sqrt{\gamma }E\frac{dE}{dz_{a}}=0,  \label{fin15} \\
& \dot{z_{a}}-N_{0}\frac{\frac{p_{a}}{\gamma }}{\sqrt{\frac{p_{a}^{2}}{%
\gamma }+m_{a}^{2}}}+\frac{N_{1}}{\gamma }=0\;.  \label{fin15a}
\end{align}

\bigskip In the equations (\ref{fin15}),(\ref{fin15a}) the prime denotes
derivative with respect to $x$ and the dot denotes derivative with respect
to $t$. All metric components ($N_{0}$, $N_{1}$, $\gamma $) are evaluated at
the point $x=z_{a}$ where%
\begin{equation}
\frac{\partial f}{\partial z_{a}}\equiv \left. \frac{\partial f(x)}{\partial
x}\right| _{x=z_{a}}\;.  \label{derivz}
\end{equation}%
Previous work in the $e_{a}=0$ case \cite{rkern} has shown that an
equilibrium solution exists when all particles have the same mass and are
evenly spaced around the circle. \ So it is reasonable to search for
(possibly similar) equilibrium solutions in the charged case. \ Introducing
the ansatz $m_{a}=m,\dot{z}_{a}=0,p_{a}=0$ into eqs. (\ref{fin-9})-(\ref%
{fin15a}) leads to the simplification

\begin{eqnarray}
2\Pi ^{\prime }-\Pi \Psi ^{\prime } &=&0;  \label{final-1} \\
\Psi ^{\prime \prime }-\frac{1}{4}(\Psi ^{\prime })^{2}-(\kappa \Pi
)^{2}+\gamma \left( \tau ^{2}-\frac{\Lambda -\kappa E^{2}}{2}\right) && 
\notag \\
+\kappa \sum_{a}\sqrt{\gamma }m\delta (x-z_{a}) &=&0  \label{final-2} \\
N_{0}^{\prime \prime }-\dot{\tau}\gamma -N_{0}\left\{ \gamma \left( \tau
^{2}-\frac{\Lambda -\kappa E^{2}}{2}\right) +\kappa /2\sum_{a}\sqrt{\gamma }%
m\delta (x-z_{a})\right\} &=&0  \label{final-3} \\
N_{1}^{\prime }-\dot{\gamma}/2+\gamma \tau N_{0} &=&0  \label{final-4} \\
\dot{\Pi}+\partial _{1}(-\frac{1}{\gamma }N_{1}\Pi +\frac{1}{2\kappa \sqrt{%
\gamma }}N_{0}\Psi ^{\prime }+\frac{1}{\kappa \sqrt{\gamma }}N_{0}^{\prime
}) &=&0  \label{final-5} \\
\dot{\Psi}+2N_{0}\left( \kappa \frac{\Pi }{\sqrt{\gamma }}+\tau \right)
-N_{1}(\frac{1}{\gamma }\Psi ^{\prime }) &=&0  \label{final-6} \\
\frac{\partial N_{0}}{\partial z_{a}}m+\int dxN_{0}\sqrt{\gamma }E\frac{dE}{%
dz_{a}} &=&0  \label{final-7} \\
N_{1}(z_{a}) &=&0\;.  \label{final-8}
\end{eqnarray}

\section{Two Neutral Particles of Unequal Mass}

We begin by attempting to solve the case of two uncharged bodies of unequal
mass. Here we find that the resulting constraint equations seem to force
very specific conditions on the system. We shall encounter a similar
situation in the charged case

As before we use the same field equations (\ref{fin-9})-(\ref{fin15a}) and
again equilibrium solutions are characterized by $\dot{z_{a}}=0=p_{a}$,
corresponding to a situation in which all particles are motionless at
various points around the circle. This implies from (\ref{fin15},\ref{fin15a}%
) that 
\begin{equation}
\frac{\partial N_{0}}{\partial z_{a}}=0\text{ \ \ \ \ \ \ \ \ \ \ }%
N_{1}(z_{a})=0\;\;.  \label{eqzeqs}
\end{equation}%
the field equations (\ref{fin-9}--\ref{fin14}) can be solved in terms of the
quantity $c^{2}\equiv \gamma \left( \tau ^{2}-\Lambda /2\right) $, which is
a function only of the time coordinate.

The solution to (\ref{fin-9}) is simply

\begin{equation}
\Pi =\Pi _{0}(t)\exp {\Psi /2}  \label{firstsoln}
\end{equation}%
which can then be substituted into (\ref{fin10}) to obtain

\begin{equation}
\Psi ^{\prime \prime }-\frac{1}{4}(\Psi ^{\prime })^{2}-(\kappa \Pi
_{0})^{2}\exp {\Psi }+c^{2}+\sum_{a=1}^{N}\kappa \sqrt{\gamma }%
\;m_{a}\;\delta (x-z_{a})=0  \label{oldeqn}
\end{equation}%
for a system of $N$ particles in equilibrium on the circle.

When $N=2$, the two particles are placed on a line at coordinates $\left(
z_{1},z_{2}\right) $ respectively, dividing the line into three regions.
Without loss of generality, a shift in the $x$-coordinate allows us to place
the origin equidistant from either mass, so that $z_{1}=z=-z_{2}$.

\bigskip In the $2$-body case the matching conditions imply from (\ref%
{oldeqn}) that%
\begin{eqnarray}
\Psi _{\text{\textsc{1}}}(L) &=&\Psi _{\text{\textsc{3}}}(-L)\text{ \ \ \ \
\ \ \ \ \ \ \ \ \ \ \ \ \ \ \ \ \ \ \ \ }\Psi _{\text{\textsc{1}}}^{\prime
}(L)=\Psi _{\text{\textsc{3}}}^{\prime }(-L)  \label{outmatch} \\
\Psi _{\text{\textsc{1}}}(z) &=&\Psi _{\text{\textsc{2}}}(z)\text{ \ \ \ \ \
\ \ \ \ \ \ \ \ \ \ \ \ \ \ \ \ \ \ \ \ \ \ }\Psi _{\text{\textsc{3}}%
}(-z)=\Psi _{\text{\textsc{2}}}(-z)  \label{contmatch} \\
\Psi _{\text{\textsc{1}}}^{\prime }(z)-\Psi _{\text{\textsc{2}}}^{\prime
}(z) &=&-\kappa \sqrt{\gamma }m_{1}\text{ \ \ \ \ \ \ }\Psi _{\text{\textsc{2%
}}}^{\prime }(-z)-\Psi _{\text{\textsc{3}}}^{\prime }(-z)=-\kappa \sqrt{%
\gamma }m_{2}  \label{massmatch}
\end{eqnarray}%
for the dilaton field. \ The conditions (\ref{outmatch}) are smoothness
conditions at the join of the outer regions where the line segment is
identified to a circle. \ Conditions (\ref{contmatch}) are continuity
conditions at the locations of the particles, and conditions (\ref{massmatch}%
) are junction conditions which follow upon integration of (\ref{oldeqn}).

In general there are three possible choices for $c^{2}$\textbf{: }$c^{2}=0,$ 
$\ \ c^{2}=c_{+}^{2}>0,$ $\ c^{2}=c_{-}^{2}<0$.\textbf{\ \ }It is
straightforward to show that no solutions exist for $c^{2}=0$. \ The
solution in each region \textsc{I=1,..,3 }when $c^{2}=0$ is 
\begin{equation}
{\Psi }_{\text{\textsc{I}}}=-2\ln \left[ \kappa \Pi _{0}\left( {B}_{\text{%
\textsc{I}}}+\left( s_{\text{\textsc{I}}}x+\left( \text{\textsc{I-2}}\right)
L\right) \right) \right]  \label{psisolc0}
\end{equation}%
where ${B}_{\text{\textsc{I}}}$ are functions only of time. The constant $%
\left( \text{\textsc{I-2}}\right) L$ is inserted for convenience, and $s_{%
\text{\textsc{I}}}^{2}=1$ for \textsc{I}=\textsc{2} and $s_{\text{\textsc{I}}%
}=1$ otherwise. \ eq. (\ref{outmatch}) \ implies that ${B}_{\text{\textsc{1}}%
}={B}_{\text{\textsc{3}}}={B}$, and \ (\ref{contmatch}) is easily shown to
imply that ${B}_{\text{\textsc{2}}}={B}$, and $z=L/2$, with $s_{\text{%
\textsc{2}}}=-1$; the only alternative solution yields $L=0$, which is
unacceptable. \ Finally, equation (\ref{massmatch}) implies after some
algebra that%
\begin{equation}
{B=}\frac{4}{\kappa \sqrt{\gamma }}\left( \frac{1}{m_{2}}-\frac{1}{m_{1}}%
\right) \text{ \ \ \ \ \ \ }L{=}\frac{-4}{\kappa \sqrt{\gamma }}\left( \frac{%
1}{m_{2}}+\frac{1}{m_{1}}\right)  \label{BLc0}
\end{equation}%
which is not allowed, since $L>0$. \ \ Hence there are no equilibrium
solutions if $c^{2}=0$, consistent with the single-particle case.

Consequently we need only examine the cases $c^{2}=c_{+}^{2}>0\ $\ and $%
c^{2}=c_{-}^{2}<0$. \ Our analysis shall concentrate primarily on the former
case, since the latter case can be obtained from it by setting $%
c\longrightarrow ic$\ in our solutions, with only slight modifications that
we shall discuss as they arise.

For the case $c^{2}=c_{+}^{2}>0$, the solution in each region \textsc{%
I=1,..,3 }is 
\begin{equation}
{\Psi }_{\text{\textsc{I}}}=-2\ln \left[ {B}_{\text{\textsc{I}}}-\cosh
\left( c_{+}\left( x+\left( \text{\textsc{I-2}}\right) L\right) +{A}_{\text{%
\textsc{I}}}\right) \right] -2\ln \left[ \frac{\kappa \Pi _{0}}{c_{+}\sqrt{{B%
}_{\text{\textsc{I}}}^{2}-1}}\right]  \label{psisolint}
\end{equation}%
where as before $\left( {A}_{\text{\textsc{I}}},{B}_{\text{\textsc{I}}%
}\right) $ are functions only of time, and the constant $\left( \text{%
\textsc{I-2}}\right) L$ is inserted for convenience. \ Smoothness at the
identification point yields from (\ref{outmatch}) that 
\begin{eqnarray}
\frac{{B}_{\text{\textsc{1}}}-\cosh \left( {A}_{\text{\textsc{1}}}\right) }{%
\sqrt{{B}_{\text{\textsc{1}}}^{2}-1}} &=&\frac{{B}_{\text{\textsc{3}}}-\cosh
\left( {A}_{\text{\textsc{3}}}\right) }{\sqrt{{B}_{\text{\textsc{3}}}^{2}-1}}
\label{match31a} \\
\frac{{\sinh }\left( {A}_{\text{\textsc{1}}}\right) }{{B}_{\text{\textsc{1}}%
}-\cosh \left( {A}_{\text{\textsc{1}}}\right) } &=&\frac{{\sinh }\left( {A}_{%
\text{\textsc{3}}}\right) }{{B}_{\text{\textsc{3}}}-\cosh \left( {A}_{\text{%
\textsc{3}}}\right) }  \label{match31b}
\end{eqnarray}%
It is straightforward to show that the only solution to these equations is ${%
A}_{\text{\textsc{1}}}={A}_{\text{\textsc{3}}}={A},{B}_{\text{\textsc{1}}}={B%
}_{\text{\textsc{3}}}={B}$. \ 

Consider next matching the solutions at the points $\left(
z_{1},z_{2}\right) $. Eqs. (\ref{contmatch}) yield%
\begin{eqnarray}
\frac{{B}-\cosh \left( c_{+}\left( z-L\right) +{A}\right) }{\sqrt{{B}^{2}-1}}
&=&\frac{{B}_{\text{\textsc{2}}}-\cosh \left( c_{+}z+{A}_{\text{\textsc{2}}%
}\right) }{\sqrt{{B}_{\text{\textsc{2}}}^{2}-1}}  \label{match12a} \\
\frac{{B}-\cosh \left( c_{+}\left( z-L\right) -{A}\right) }{\sqrt{{B}^{2}-1}}
&=&\frac{{B}_{\text{\textsc{2}}}-\cosh \left( c_{+}z-{A}_{\text{\textsc{2}}%
}\right) }{\sqrt{{B}_{\text{\textsc{2}}}^{2}-1}}  \label{match12b}
\end{eqnarray}%
whereas%
\begin{eqnarray}
\frac{{2}c_{+}\sinh \left( c_{+}\left( z-L\right) +{A}\right) }{{B}-\cosh
\left( c_{+}\left( z-L\right) +{A}\right) }-\frac{{2}c_{+}\sinh \left(
c_{+}z+{A}_{\text{\textsc{2}}}\right) }{{B}_{\text{\textsc{2}}}-\cosh \left(
c_{+}z+{A}_{\text{\textsc{2}}}\right) } &=&-\kappa \sqrt{\gamma }m_{1}
\label{match12c} \\
\frac{{2}c_{+}\sinh \left( c_{+}\left( z-L\right) -{A}\right) }{{B}-\cosh
\left( c_{+}\left( z-L\right) -{A}\right) }-\frac{{2}c_{+}\sinh \left(
c_{+}z-{A}_{\text{\textsc{2}}}\right) }{{B}_{\text{\textsc{2}}}-\cosh \left(
c_{+}z-{A}_{\text{\textsc{2}}}\right) } &=&-\kappa \sqrt{\gamma }m_{2}
\label{match12d}
\end{eqnarray}%
follow from eqs. (\ref{massmatch}). \ 

Eqs. (\ref{match12a}--\ref{match12d}) constitute a set of four equations for
the four unknowns ${A,{B},A_{\text{\textsc{2}}},}$ and ${B}_{\text{\textsc{2}%
}}$. \ \ Although their general solution is quite complicated, the ansatz%
\begin{equation}
{B}_{\text{\textsc{2}}}={B}  \label{ABansatz}
\end{equation}%
immediately solves (\ref{match12a},\ref{match12b}) as long as 
\begin{equation}
c_{+}\left( z-L\right) +{A=\pm }\left( c_{+}z+{A}_{\text{\textsc{2}}}\right) 
\text{ \ \ and \ \ \ \ }c_{+}\left( z-L\right) -{A=\pm }\left( c_{+}z+{A}_{%
\text{\textsc{2}}}\right) \text{\ }  \label{ABansatz2}
\end{equation}%
\noindent From eqs. (\ref{match12c},\ref{match12d}), we find that 
\begin{equation}
{A_{\text{\textsc{2}}}=-A}\text{ \ \ \ \ \ \ and \ \ \ \ \ }z=L/2
\label{ABansatz3}
\end{equation}%
or else $m_{1}=m_{2}=0$. \ After some algebra we obtain 
\begin{eqnarray}
{B} &{=}&\left\{ \frac{2c_{+}}{\kappa \sqrt{\gamma }}\left\{ \frac{%
m_{2}-m_{1}}{m_{1}m_{2}}{\cosh \left( \frac{c_{+}L}{2}\right) \sinh \alpha }%
\right\} \right.  \notag \\
&&\left. {+}\left[ \frac{m_{2}+m_{1}}{m_{1}m_{2}}\sinh \left( \frac{c_{+}L}{2%
}\right) +\frac{\kappa \sqrt{\gamma }}{2c_{+}}\cosh \left( \frac{c_{+}L}{2}%
\right) \right] {\cosh \alpha }\right\}  \notag \\
&{\equiv }&\beta  \label{Bsoln2a} \\
{A} &{=}&{\tanh }^{-1}\left[ \frac{\frac{m_{2}-m_{1}}{m_{1}m_{2}}{\sinh
\left( \frac{c_{+}L}{2}\right) }}{\frac{m_{2}+m_{1}}{m_{1}m_{2}}\cosh \left( 
\frac{c_{+}L}{2}\right) +\frac{\kappa \sqrt{\gamma }}{2c_{+}}\sinh \left( 
\frac{c_{+}L}{2}\right) }\right] {\equiv \alpha }  \label{Asoln2}
\end{eqnarray}%
as the solution to (\ref{match12c},\ref{match12d}). \ 

This yields \ 
\begin{equation}
{\Psi }=-2\ln \left[ {\beta }-\cosh \left( c_{+}f(x)+\alpha \right) \right]
-2\ln \left[ \frac{\kappa \Pi _{0}}{c_{+}\sqrt{{\beta }^{2}-1}}\right]
\label{psi2sol}
\end{equation}%
for the dilaton, where 
\begin{equation}
f\left( x\right) =\left\{ 
\begin{array}{c}
x-L\text{ \ \ \ \ \ \ \ \ \ \ \ \ \ \ \ \ }x>L/2 \\ 
-x\text{\ \ \ \ \ \ \ \ \ \ }L/2>x>-L/2 \\ 
x+L\text{ \ \ \ \ \ \ \ \ \ \ }-L/2>x%
\end{array}%
\right.  \label{saw2}
\end{equation}%
is the continuous sawtooth function, with kinks at the locations of the
particles. Substitution of (\ref{psi2sol}) into (\ref{firstsoln}) yields 
\begin{equation}
\kappa \Pi =\frac{\pm c_{+}\sqrt{\beta ^{2}-1}}{\beta -\cosh
(c_{+}f(x)+\alpha )}  \label{Pi2sol}
\end{equation}%
as the solution for $\Pi $.

Consider next eq. (\ref{fin11}), which can be written as 
\begin{equation}
N_{0}^{\prime \prime }=\dot{\tau}\gamma +N_{0}\left\{ c_{+}^{2}+\kappa
/2\sum_{a}\sqrt{\gamma }\;m_{a}\;\delta (x-z_{a}(x^{0}))\right\}
\label{fin112}
\end{equation}%
Proceeding as before, the general solution is 
\begin{equation}
N_{0\text{\textsc{I}}}=-\frac{\dot{\tau}\gamma }{c_{+}^{2}}+\lambda _{\text{%
\textsc{I}}}\cosh \left( c_{+}\left( x+\left( \text{\textsc{I-2}}\right)
L\right) +{C}_{\text{\textsc{I}}}\right)  \label{Nsolint}
\end{equation}%
The smoothness conditions at the join are the same as in eqs. (\ref{outmatch}%
) , with $\Psi $ replaced by $N_{0}$. These imply $\lambda _{\text{\textsc{1}%
}}=\lambda _{\text{\textsc{3}}}$ and ${C}_{\text{\textsc{1}}}={C}_{\text{%
\textsc{3}}}={C}$. \ The same is true for conditions (\ref{contmatch}),
which then yield $\lambda _{\text{\textsc{1}}}=\lambda _{\text{\textsc{2}}}$
and ${C}_{\text{\textsc{2}}}=-{C}$. The junction conditions (\ref{massmatch}%
), however, are replaced by the conditions 
\begin{equation}
N_{0\text{\textsc{1}}}^{\prime }(z)-N_{0\text{\textsc{2}}}^{\prime }(z)=%
\frac{\kappa \sqrt{\gamma }}{2}m_{1}N_{0\text{\textsc{1}}}(z)\text{\ \ \ \ \
\ }N_{0\text{\textsc{2}}}^{\prime }(-z)-N_{0\text{\textsc{3}}}^{\prime }(-z)=%
\frac{\kappa \sqrt{\gamma }}{2}m_{2}N_{0\text{\textsc{2}}}(-z)
\label{massN0match}
\end{equation}%
where $z=L/2$. \ After addition and subtraction we find 
\begin{equation}
\lambda _{\text{\textsc{I}}}=\frac{\dot{\tau}\gamma }{\beta c_{+}^{2}}\text{
\ \ \ \ \ \ }{C=\alpha }\text{\ }  \label{betasol}
\end{equation}%
so that 
\begin{equation}
N_{0}=\frac{\dot{\tau}\gamma }{\beta c_{+}^{2}}\left( \cosh \left(
c_{+}f(x)+\alpha \right) -\beta \right)  \label{N02sol}
\end{equation}%
where $f(x)$ is the sawtooth function given in (\ref{saw2}). \ \ 

Integration of (\ref{fin12}) yields 
\begin{equation}
N_{1\text{\textsc{I}}}=\gamma \frac{\dot{c_{+}}}{c_{+}}\tilde{f}(x)-\frac{%
\gamma ^{2}\tau \dot{\tau}}{c_{+}^{3}\beta }\sinh \left( c_{+}\tilde{f}%
(x)+f^{\prime }(x)\alpha \right) +K_{\text{\textsc{I}}}  \label{N12sol}
\end{equation}%
where the $K_{\text{\textsc{I}}}$ are constants of integration and 
\begin{equation}
\tilde{f}\left( x\right) =\left\{ 
\begin{array}{c}
x-L\text{\ \ \ \ \ \ \ \ \ \ \ \ \ \ \ \ \ \ }x>L/2 \\ 
x\text{\ \ \ \ \ \ \ \ \ \ \ }L/2>x>-L/2 \\ 
x+L\text{\ \ \ \ \ \ \ \ \ \ \ \ \ \ }-L/2>x%
\end{array}%
\right.  \label{saw3}
\end{equation}%
is the discontinuous sawtooth function, with discontinuities at the
locations $x=\pm \frac{L}{2}$ of the particles. \ However the function $%
N_{1}\left( x,t\right) $ must be continuous at these locations and smooth at 
$x=\pm L$, which implies that 
\begin{subequations}
\begin{eqnarray}
\beta c_{+}^{2}\dot{c_{+}}L-\gamma \tau \dot{\tau}\left[ \sinh (\frac{c_{+}L%
}{2}+\alpha )+\sinh (\frac{c_{+}L}{2}-\alpha )\right] &=&0  \label{beta2eq}
\\
K_{\text{\textsc{1}}}=K_{\text{\textsc{3}}}=-K_{\text{\textsc{2}}}=\frac{%
\gamma ^{2}\tau \dot{\tau}}{2c_{+}^{3}\beta }\left[ \sinh (\frac{c_{+}L}{2}%
+\alpha )-\sinh (\frac{c_{+}L}{2}-\alpha )\right] &&  \label{K2eq}
\end{eqnarray}%
where the latter condition holds since $N_{1}\left( x,t\right) $ must vanish
at the locations $z_{a}$ of the particles from eq. (\ref{fin15a}).

\bigskip

Employing the definition of $\beta $ from (\ref{Bsoln2a}) we find from (\ref%
{beta2eq}) 
\end{subequations}
\begin{eqnarray}
\frac{m_{2}-m_{1}}{m_{1}m_{2}}{\sinh \alpha \frac{d}{d\tau }\sinh \left( 
\frac{c_{+}L}{2}\right) } &&  \notag \\
{+\cosh \alpha }\frac{d}{d\tau }\left[ \frac{m_{2}+m_{1}}{m_{1}m_{2}}\cosh
\left( \frac{c_{+}L}{2}\right) +\frac{\kappa \sqrt{\gamma }}{2c_{+}}\sinh
\left( \frac{c_{+}L}{2}\right) \right] &=&0
\end{eqnarray}%
which from (\ref{Asoln2}) integrates to%
\begin{equation}
\left[ \cosh \left( \frac{c_{+}L}{2}\right) +\frac{\kappa M\sqrt{\gamma }}{%
2c_{+}}\sinh \left( \frac{c_{+}L}{2}\right) \right] ^{2}-\Delta ^{2}\sinh
^{2}\left( \frac{c_{+}L}{2}\right) =\xi ^{2}  \label{gameq2}
\end{equation}%
where%
\begin{equation}
M=\frac{m_{1}m_{2}}{m_{2}+m_{1}}\text{ \ \ \ \ \ \ \ \ }\Delta =\frac{%
m_{2}-m_{1}}{m_{2}+m_{1}}  \label{2bdparams}
\end{equation}%
and $\xi $ is a constant of integration. \ 

\bigskip Inverting this equation to find $\sqrt{\gamma }$ as a function of $%
\tau $, we find 
\begin{equation}
\sqrt{\gamma }=\frac{2}{L\sqrt{\tau ^{2}-\Lambda /2}}\text{arctanh}\left( 
\frac{\xi \sqrt{\frac{\kappa ^{2}M^{2}}{4}+(\xi ^{2}-1)\left( 1-\Delta
^{2}/\xi ^{2}\right) (\tau ^{2}-\Lambda /2)}-\frac{\kappa M}{2}}{\frac{%
\kappa ^{2}M^{2}}{4}+\xi ^{2}\left( 1-\Delta ^{2}/\xi ^{2}\right) (\tau
^{2}-\Lambda /2)}\sqrt{\tau ^{2}-\Lambda /2}\right)  \label{gam2result}
\end{equation}%
and subsequently%
\begin{eqnarray*}
\alpha &=&\arctan \text{h}\left[ \Delta \frac{\frac{\kappa M}{2}\xi ^{2}-\xi 
\sqrt{\frac{\kappa ^{2}m^{2}}{4}+(\xi ^{2}-1)\left( 1-\Delta ^{2}/\xi
^{2}\right) (\tau ^{2}-\Lambda /2)}}{\left( \Delta ^{2}-\xi ^{2}\right)
(\tau ^{2}-\Lambda /2)+\frac{\kappa ^{2}m^{2}}{4}\xi ^{2}}\sqrt{(\tau
^{2}-\Lambda /2)}\right] \\
\beta &=&\frac{2}{\kappa M\xi }\left( \left( \Delta -1\right) ^{2}(\tau
^{2}-\Lambda /2)-\frac{\kappa ^{2}m^{2}}{4}\right) ^{-1}\left( \left( \Delta
+1\right) ^{2}(\tau ^{2}-\Lambda /2)-\frac{\kappa ^{2}m^{2}}{4}\right) ^{-1}
\\
&&\left[ \kappa M\Delta ^{2}(\tau ^{2}-\Lambda /2)\left( \left( \Delta
^{2}+1-2\xi ^{2}\right) (\tau ^{2}-\Lambda /2)-\frac{\kappa ^{2}m^{2}}{4}%
\right) \right. \\
&&-\xi \left( \left( \Delta ^{2}+1\right) (\tau ^{2}-\Lambda /2)-\frac{%
\kappa ^{2}m^{2}}{4}\right) \left( \left( \Delta ^{2}-1\right) (\tau
^{2}-\Lambda /2)+\frac{\kappa ^{2}m^{2}}{4}\right) \\
&&\left. \times \sqrt{\frac{\kappa ^{2}m^{2}}{4}+(\xi ^{2}-1)\left( 1-\Delta
^{2}/\xi ^{2}\right) (\tau ^{2}-\Lambda /2)}\right]
\end{eqnarray*}%
from eqs. (\ref{Bsoln2a},\ref{Asoln2}).

\bigskip eq. (\ref{fin13}) can be simplified to 
\begin{equation}
\left[ \ln (\kappa \Pi )\right] ^{\cdot }-\frac{N_{1}^{\prime }}{\gamma }-%
\frac{N_{1}}{\gamma }[\ln (\kappa \Pi )]^{\prime }=0  \label{Pi2last}
\end{equation}%
since $N_{0}\exp {\Psi /2}$ is independent of $x$. Insertion of (\ref{Pi2sol}%
) and (\ref{N12sol}) into the left-hand side (\ref{Pi2last}) yields 
\begin{eqnarray}
\frac{1}{\cosh \left( c_{+}x-\alpha \right) -\beta }\left[ \frac{1}{\beta }%
\left( \frac{\beta \dot{\beta}}{\beta ^{2}-1}-\frac{\gamma ^{2}\tau \dot{\tau%
}}{c_{+}^{2}}\right) \left( \beta \cosh \left( c_{+}x-\alpha \right)
-1\right) \right. &&  \label{Pibad} \\
-\left. \sinh \left( c_{+}x-\alpha \right) \left( \dot{\alpha}-\frac{\gamma
^{2}\tau \dot{\tau}}{c_{+}^{2}}\cosh \left( \frac{c_{+}L}{2}\right) \sinh
\left( \alpha \right) \right) \right] &=&0  \notag
\end{eqnarray}%
in the region $\left| x\right| <L/2$.

The preceding equation (\ref{Pibad}) has no solutions unless $\Delta =0$.
The problem is that each term has to vanish separately, ie. 
\begin{equation*}
\left( \frac{\beta \dot{\beta}}{\beta ^{2}-1}-\frac{\gamma ^{2}\tau \dot{\tau%
}}{c_{+}^{2}}\right) =0\text{ \ \ and \ \ }\dot{\alpha}-\frac{\gamma
^{2}\tau \dot{\tau}}{c_{+}^{2}}\cosh \left( \frac{c_{+}L}{2}\right) \sinh
\left( \alpha \right) =0
\end{equation*}%
and this is not allowed from the definitions of \ $\alpha $\ and $\beta $.
This peculiar situation suggests (within the context of our ansatz) that
only the equal mass case is allowed as a solution to the field equations due
to the consistent time-evolution of $\Pi $. \ It is remarkable that the
unequal mass case can be exactly integrated, yet the $\dot{\Pi}$ equation
cannot be solved. \ When this calculation is repeated for the $%
c^{2}=c_{-}^{2}<0$ case the same result occurs: the field equations can only
be solved when the particles have equal mass.

\section{Two-Particle Case}

Because the total charge of the system must be zero there are no solutions
for a single charged particle. \ We consider in this section the
two-particle case, which is the simplest non-trivial charged circular
self-gravitating system. \ 

Since there is freedom in choosing the origin we choose the origin to be
halfway between the two particles, which are \ thus are located at $z$ and $%
-z$. Since the total charge is zero we can assume (without loss of
generality)\ that the particle at $-z$ has charge $q$ and the particle at $z$
has charge $-q$. \ Therefore the electric field is given by the equation $E=%
\frac{1}{2}q(\epsilon (x+z)-\epsilon (x-z))$ $+E_{0}$ in this case\ (and
consequently $E^{2}=(\frac{1}{2}q^{2}+qE_{0})(\epsilon (x+z)-\epsilon
(x-z))+E_{0}^{2}$). \ 

We next consider solving the field equations (\ref{final-1})-(\ref{final-8}%
). \ The solution to equation (\ref{final-1}) yields the result 
\begin{equation}
\Pi =\Pi _{0}(t)\exp (\Psi /2),  \label{Pi}
\end{equation}%
where $\Pi _{0}(t)$ is an arbitrary function. \ It is useful at this stage
to introduce some new variables

\begin{eqnarray}
&&\Lambda _{out}=\Lambda -\kappa E_{0}^{2}, \\
&&\Lambda _{in}=\Lambda -\kappa (q^{2}+2E_{0}q+E_{0}^{2})
\end{eqnarray}%
so that in equations (\ref{final-2}) and (\ref{final-3}) the term $\kappa
E^{2}$ is essentially just an adjustment to the value of the cosmological
constant $\Lambda $. \ This is similar to the result of (1+1) lineal gravity
with charge in which the electric field becomes sort of a ``local''
cosmological constant. \ By introducing $\Lambda _{out}$ and $\Lambda _{in}$
it is possible to write%
\begin{equation}
\gamma (\tau ^{2}-\Lambda /2+\kappa E^{2}/2)=\left\{ 
\begin{array}{c}
\gamma (\tau ^{2}-\Lambda _{out}/2)\text{ }\equiv c^{2\text{ }}\text{when }%
|x|>z \\ 
\gamma (\tau ^{2}-\Lambda _{in}/2)\text{ }\equiv d^{2}\text{ \ when }|x|<z%
\end{array}%
\right\}  \label{def-c}
\end{equation}%
where $x$ is periodic with period $2L$. \ 

We shall assume that $c^{2}=c_{+}^{2}>0,d^{2}=d_{+}^{2}>0$. The analysis for 
$c^{2}=c_{-}^{2}<0,d^{2}=d_{-}^{2}<0$ is fully analogous and we shall only
discuss the results for this alternate case\textbf{. }

\subsection{Solving for $N_{0}$}

In order to solve for $N_{0}$ we will first integrate (\ref{final-3}) across
the particles to get the following two matching conditions across the
particles for $N_{0}^{\prime }$: 
\begin{eqnarray}
&&N_{0}^{\prime }(z+\epsilon )-N_{0}^{\prime }(z-\epsilon )=\frac{\kappa m%
\sqrt{\gamma }}{2}N_{0}(z), \\
&&N_{0}^{\prime }(-z+\epsilon )-N_{0}^{\prime }(-z-\epsilon )=\frac{\kappa m%
\sqrt{\gamma }}{2}N_{0}(-z)  \label{N0-con2}
\end{eqnarray}%
\qquad

Then we also assume that $N_{0}$ should be a continuous function, which
gives the following matching conditions%
\begin{eqnarray}
&&N_{0}(z-\epsilon )=N_{0}(z+\epsilon ), \\
&&N_{0}(-z-\epsilon )=N_{0}(-z+\epsilon )
\end{eqnarray}%
where the limit $\epsilon \rightarrow 0$ is taken throughout.

Solving (\ref{final-3}) individually in the regions between the particles is
straight-forward and leads to a general solution for $N_{0}$

\begin{equation}
N_{0}=\left\{ 
\begin{array}{c}
A\cosh (d_{+}x)+E\sinh (d_{+}x)-\frac{\dot{\tau}\gamma }{d_{+}^{2}},|x|<z \\ 
C\sinh (c_{+}(L-|x|))+D\cosh (c_{+}(L-|x|))-\frac{\dot{\tau}\gamma }{%
c_{+}^{2}},|x|>z%
\end{array}%
\right\}  \label{N0-sol}
\end{equation}%
, where $A(t),E(t),C(t)$ and $D(t)$ are $x$-independent. \ We can then use
the four matching conditions across the particles and the smoothness
conditions to solve for the four unknowns $A,E,C,D$:. \ 
\begin{eqnarray}
E &=&C=0 \\
D &=&A\frac{\cosh (d_{+}z)}{\cosh (c_{+}(z-L))}+\frac{\dot{\tau}\gamma }{%
\cosh (c_{+}(z-L))}\left( \frac{1}{c_{+}^{2}}-\frac{1}{d_{+}^{2}}\right) \\
A &=&\left( \frac{\dot{\tau}\gamma }{Bd_{+}^{2}}\right) \left( \frac{\kappa m%
\sqrt{\gamma }}{2}+c_{+}d_{+}^{2}\tanh (c_{+}(z-L))\left( \frac{1}{c_{+}^{2}}%
-\frac{1}{d_{+}^{2}}\right) \right)
\end{eqnarray}%
where%
\begin{equation}
B=\frac{\kappa m\sqrt{\gamma }}{2}\cosh (d_{+}z)+d_{+}\sinh
(d_{+}z)-c_{+}\cosh (d_{+}z)\tanh (c_{+}(z-L))
\end{equation}%
or we can write $D$ in the form%
\begin{eqnarray}
D &=&\frac{\kappa m\sqrt{\gamma }}{2}\left( \frac{\dot{\tau}\gamma }{%
Bd_{+}^{2}}\right) \frac{\cosh (d_{+}z)}{\cosh (c_{+}(z-L))} \\
&&+\left( \frac{c_{+}}{B}\tanh (c_{+}(z-L))\cosh (d_{+}z)+1\right) \frac{%
\dot{\tau}\gamma }{\cosh (c_{+}(z-L))}\left( \frac{1}{c_{+}^{2}}-\frac{1}{%
d_{+}^{2}}\right)  \notag
\end{eqnarray}

Note the resulting solution also has the following the symmetry properties 
\begin{eqnarray*}
N_{0}(-z+x) &=&N_{0}(z-x),\text{ \ \ \ \ \ }N_{0}(-z-x)=N_{0}(z+x) \\
N_{0}^{\prime }(-z+x) &=&-N_{0}^{\prime }(z-x),\text{ \ \ }N_{0}^{\prime
}(-z-x)=-N_{0}^{\prime }(z+x)
\end{eqnarray*}%
\ 

\subsection{Solving for $\Psi $}

To obtain the solution for $\Psi $ we make use of an ansatz previously
employed in the neutral $N$-body case \cite{rkern}\ 
\begin{equation}
\Psi =-2\ln (-N_{0}(x)\frac{\beta h^{2}}{\dot{\tau}\gamma })-2\ln \left( 
\frac{\kappa \Pi _{0}}{h\sqrt{\beta ^{2}-1}}\right)
\end{equation}%
where 
\begin{equation}
h\equiv \left\{ 
\begin{array}{c}
c_{+},|x|>z \\ 
d_{+},|x|<z%
\end{array}%
\right\}
\end{equation}%
\begin{equation}
\beta \equiv \left\{ 
\begin{array}{c}
\beta _{out},|x|>z \\ 
\beta _{in},|x|<z%
\end{array}%
\right\}
\end{equation}

Note that $\beta _{out}$ and $\beta _{in}$ are constants in terms of $x$. \
Using this ansatz along with equations (\ref{final-3}) and (\ref{Pi}) we
find 
\begin{eqnarray*}
\Psi ^{\prime } &=&-2\frac{N_{0}^{\prime }}{N_{0}} \\
\Psi ^{\prime \prime } &=&-2\frac{N_{0}^{\prime \prime }}{N_{0}}+2\left( 
\frac{N_{0}^{\prime }}{N_{0}}\right) ^{2}=-\frac{2\dot{\tau}\gamma }{N_{0}}%
+2\left( \frac{N_{0}^{\prime }}{N_{0}}\right) ^{2}-2h^{2}-\kappa \sum_{a}%
\sqrt{\gamma }m\;\delta (x-z_{a}) \\
(\kappa \Pi )^{2} &=&(\kappa \Pi _{0})^{2}\exp (\Psi )=\frac{(\dot{\tau}%
\gamma )^{2}(\beta ^{2}-1)}{N_{0}^{2}\beta ^{2}h^{2}}
\end{eqnarray*}%
Inserting these results into (\ref{final-2}) leads to the equation%
\begin{equation*}
(N_{0}^{\prime })^{2}-2\dot{\tau}\gamma N_{0}-h^{2}N_{0}^{2}-\frac{(\dot{\tau%
}\gamma )^{2}(\beta ^{2}-1)}{\beta ^{2}h^{2}}=0
\end{equation*}%
and substituting the equation for $N_{0}$ into each region individually
allows the calculation of $\beta _{out}$ and $\beta _{in}$. \ For example
when $|x|>z$ 
\begin{align}
D^{2}c_{+}^{2}(\sinh ^{2}(c_{+}(L-|x|))-\cosh ^{2}(c_{+}(L-|x|)))& =-\frac{(%
\dot{\tau}\gamma )^{2}}{c_{+}^{2}\beta _{out}^{2}}  \notag \\
\text{ or alternatively }\beta _{out}& =\frac{\dot{\tau}\gamma }{Dc_{+}^{2}}
\label{beta-out}
\end{align}%
A similar calculation for $|x|<z$ yields 
\begin{equation}
\beta _{in}=\frac{\dot{\tau}\gamma }{Ad_{+}^{2}}  \label{beta-in}
\end{equation}

To summarize, the solution for (\ref{final-2}) is given by the equation

\begin{equation}
\Psi =\left\{ 
\begin{array}{c}
-2\ln (-N_{0}(x))-2\ln \left( \frac{\kappa \Pi _{0}c_{+}\beta _{out}}{\dot{%
\tau}\gamma \sqrt{\beta _{out}^{2}-1}}\right) ,|x|>z \\ 
-2\ln (-N_{0}(x))-2\ln \left( \frac{\kappa \Pi _{0}d_{+}\beta _{in}}{\dot{%
\tau}\gamma \sqrt{\beta _{in}^{2}-1}}\right) ,|x|<z%
\end{array}%
\right\}  \label{psi-sol}
\end{equation}

where $\beta _{out}$ and $\beta _{in}$ are respectively given by eqs. (\ref%
{beta-out}) and (\ref{beta-in}).

\subsection{Solving for $N_{1}$}

Integrating (\ref{fin12}) yields the general solution for $N_{1}$

\begin{equation}
N_{1}=\left\{ 
\begin{array}{c}
\left( \frac{\dot{\gamma}}{2}+\frac{\gamma ^{2}\tau \dot{\tau}}{c_{+}^{2}}%
\right) x-\frac{\gamma \tau D}{c_{+}}\sinh (c_{+}(x+L))+K_{1},x<-z \\ 
\left( \frac{\dot{\gamma}}{2}+\frac{\gamma ^{2}\tau \dot{\tau}}{d_{+}^{2}}%
\right) x-\frac{\gamma \tau A}{d_{+}}\sinh (d_{+}x)+K_{2},-z<x<z \\ 
\left( \frac{\dot{\gamma}}{2}+\frac{\gamma ^{2}\tau \dot{\tau}}{c_{+}^{2}}%
\right) x-\frac{\gamma \tau D}{c_{+}}\sinh (c_{+}(x-L))+K_{3},x>z%
\end{array}%
\right\}  \label{N1prelim}
\end{equation}

To find the unknowns $K_{1},K_{2},$ and $K_{3}$ in eq. (\ref{N1prelim}), it
is convenient to use the constraint equations (\ref{final-8})\ for $N_{1}$ .
\ The constraint $N_{1}(z_{a})=0$ implies that 
\begin{eqnarray*}
N_{1}(-z-\epsilon ) &=&0,\text{ \ \ }N_{1}(-z+\epsilon )=0 \\
N_{1}(z-\epsilon ) &=&0,\text{ \ \ }N_{1}(z+\epsilon )=0
\end{eqnarray*}%
Together $N_{1}(-z+\epsilon )=0$ and $N_{1}(z-\epsilon )=0$ imply that $%
K_{2}=0$ and

\begin{equation}
\gamma \left( \frac{\dot{d}_{+}}{d_{+}}z-\frac{\tau A}{d_{+}}\sinh
(d_{+}z)\right) =0  \label{constraint1}
\end{equation}%
where we have employed the following identities

\begin{equation}
\gamma \frac{\dot{c}_{+}}{c_{+}}=\left( \frac{\dot{\gamma}}{2}+\frac{\gamma
^{2}\tau \dot{\tau}}{c_{+}^{2}}\right) ,\text{ \ \ \ \ \ \ \ \ }\gamma \frac{%
\dot{d}_{+}}{d_{+}}=\left( \frac{\dot{\gamma}}{2}+\frac{\gamma ^{2}\tau \dot{%
\tau}}{d_{+}^{2}}\right)  \label{ident-c}
\end{equation}%
that follow from eq. (\ref{def-c}).

Equation (\ref{constraint1}) is a new relation between the functions $%
d_{+},\gamma $ and $A$. Using $N_{1}(-z-\epsilon )=0$ and $N_{1}(z+\epsilon
)=0$ along with the condition $N_{1}(-L)=N_{1}(L)$ yields

\begin{eqnarray}
K_{1} &=&-K_{3}=\gamma \frac{\dot{c}_{+}}{c_{+}}L \\
\gamma \left( \frac{\dot{c}_{+}}{c_{+}}(z-L)-\frac{\tau D}{c_{+}}\sinh
(c_{+}(z-L))\right) &=&0  \label{constraint2}
\end{eqnarray}%
where (\ref{constraint2}) is another new relation between the various
functions. \ So the solution for $N_{1}$ is%
\begin{equation}
N_{1}=\left\{ 
\begin{array}{c}
\gamma \frac{\dot{c}_{+}}{c_{+}}(x+L)-\frac{\gamma \tau D}{c_{+}}\sinh
(c_{+}(x+L)),x<-z \\ 
\gamma \frac{\dot{d}_{+}}{d_{+}}x-\frac{\gamma \tau A}{d_{+}}\sinh
(d_{+}x),-z<x<z \\ 
\gamma \frac{\dot{c}_{+}}{c_{+}}(x-L)-\frac{\gamma \tau D}{c_{+}}\sinh
(c_{+}(x-L)),x>z%
\end{array}%
\right\}  \label{N1-sol}
\end{equation}%
with (\ref{final-8}) now being represented by the constraints%
\begin{eqnarray}
\frac{\dot{d}_{+}}{d_{+}}z-\frac{\tau A}{d_{+}}\sinh (d_{+}z) &=&0
\label{constraint-3} \\
\frac{\dot{c}_{+}}{c_{+}}(z-L)-\frac{\tau D}{c_{+}}\sinh (c_{+}(z-L)) &=&0
\label{constraint-4}
\end{eqnarray}

\subsection{Constraint equations}

We must now verify that the constraint equations (\ref{final-7}), (\ref%
{constraint-3}),(\ref{constraint-4}) can all be satisfied. \ It is useful to
introduce a new parameter

\begin{equation}
n\equiv \frac{\Lambda _{out}-\Lambda _{in}}{2}=\frac{\kappa q}{2}(q+2E_{0})
\label{n}
\end{equation}%
so that $d_{+}^{2}=c_{+}^{2}+\gamma n$, or 
\begin{equation}
\frac{1}{c_{+}^{2}}-\frac{1}{d_{+}^{2}}=\frac{\gamma n}{c_{+}^{2}d_{+}^{2}}
\end{equation}%
The constraint (\ref{final-7}) can then be written as 
\begin{eqnarray}
0 &=&d_{+}\tanh (d_{+}z)\frac{2\sqrt{\gamma }}{m\kappa }\left[ n+(\frac{%
\kappa m}{2})^{2}\right] +2\frac{\gamma n}{c_{+}}d_{+}\tanh (d_{+}z)\tanh
(c_{+}(z-L))  \notag \\
&&+c_{+}\tanh (c_{+}(z-L))\left[ \frac{2\gamma ^{\frac{3}{2}}n}{m\kappa
c_{+}^{2}}\left( (\frac{\kappa m}{2})^{2}-n\right) +\frac{2\sqrt{\gamma }}{%
m\kappa }\left( (\frac{\kappa m}{2})^{2}-n\right) \right]
\label{finalconstraint1}
\end{eqnarray}%
where the calculational details are given in the appendix. Rather than first
attempting to solve this equation in full generality, we instead consider
specific values of $n$.

The simplest case to examine is when $n=0$, or $q=-2E_{0}$. eq. (\ref%
{finalconstraint1}) reduces to%
\begin{equation*}
0=c_{+}\frac{\sqrt{\gamma }\kappa m}{2}\left( \tanh (c_{+}z)+\tanh
(c_{+}(z-L))\right)
\end{equation*}%
which in turn implies that $z=\frac{L}{2}$, and so\ the particles must be
located at $\frac{L}{2}$ and $-\frac{L}{2}$. \ The electric field is
therefore given by \ $E=\frac{1}{2}q(\epsilon (x+z)-\epsilon (x-z)-$ $1)$ or
more explicitly%
\begin{equation*}
E= \LARGE{  {-\frac{1}{2}q,|x|>z}{\frac{1}{2}q,|x|<z} \LARGE}
\end{equation*}%
Note that $E^{2}=\frac{q^{2}}{4}$, which is constant, so the magnitude of
the field doesn't change with position. \ 

We next rewrite \ (\ref{constraint-3}) and (\ref{constraint-4}) in a more
convenient form. \ After some manipulation, (\ref{constraint-3}) can be
written as 
\begin{eqnarray}
&&d_{+}^{2}\dot{d}_{+}z\left( \frac{\kappa m\sqrt{\gamma }}{2}\cosh
(d_{+}z)+d_{+}\sinh (d_{+}z)-c_{+}\cosh (d_{+}z)\tanh (c_{+}(z-L))\right) 
\notag \\
- &&\gamma \tau \dot{\tau}\left( \frac{\kappa m\sqrt{\gamma }}{2}+\frac{%
\gamma n}{c_{+}}\tanh (c_{+}(z-L))\right) \sinh (d_{+}z)=0
\label{finalconstraint-2}
\end{eqnarray}%
and (\ref{constraint-4}) can be written as%
\begin{eqnarray}
&&\dot{c}_{+}(z-L)-\tau \left( \frac{\dot{\tau}\gamma }{Bd_{+}^{2}}\right)
\left( \frac{\kappa m\sqrt{\gamma }}{2}+\frac{\gamma n}{c_{+}}\tanh
(c_{+}(z-L))\right) \cosh (d_{+}z)\tanh (c_{+}(z-L))  \notag \\
- &&\tau \dot{\tau}\gamma \left( \frac{\gamma n}{c_{+}^{2}d_{+}^{2}}\right)
\tanh (c_{+}(z-L))=0  \label{finalconstraint-3}
\end{eqnarray}

\bigskip

Since $n=0$, $c_{+}=d_{+}$. Recalling that $z=\frac{L}{2}$, we have 
\begin{equation}
c_{+}^{2}\dot{c}_{+}\frac{L}{2}\left( \frac{\kappa m\sqrt{\gamma }}{2}\cosh
(c_{+}\frac{L}{2})+2c_{+}\sinh (c_{+}\frac{L}{2})\right) -\gamma \tau \dot{%
\tau}\frac{\kappa m\sqrt{\gamma }}{2}\sinh (c_{+}\frac{L}{2})=0
\label{n0constraint}
\end{equation}%
for (\ref{finalconstraint-2}). It is straightforward to show that (\ref%
{finalconstraint-3}) reduces to the same equation. eq. (\ref{n0constraint})
can be integrated to yield \ 
\begin{equation*}
\cosh (\sqrt{\gamma \left( \tau ^{2}-\Lambda _{out}/2\right) }\frac{L}{2})+%
\frac{\kappa m\sqrt{\gamma }}{4c_{+}}\sinh (\sqrt{\gamma \left( \tau
^{2}-\Lambda _{out}/2\right) }\frac{L}{2})=\xi
\end{equation*}%
where $\xi $ is a constant of integration. \ It is then possible to solve
for $\sqrt{\gamma }$ 
\begin{equation}
\sqrt{\gamma }=\frac{2}{L\sqrt{\tau ^{2}-\Lambda _{out}/2}}\arctan h\left( 
\frac{\xi \sqrt{\frac{\kappa ^{2}m^{2}}{64}+(\xi ^{2}-1)\left( \tau
^{2}-\Lambda _{out}/2\right) }-\frac{\kappa m}{8}}{\frac{\kappa ^{2}m^{2}}{64%
}+\xi ^{2}\left( \tau ^{2}-\Lambda _{out}/2\right) }\sqrt{\left( \tau
^{2}-\Lambda _{out}/2\right) }\right)  \label{rtgam}
\end{equation}

\bigskip

One might expect that a solution of (\ref{finalconstraint1}) \ for general $%
n $ would provide information on the locations of the particles. However due
to the extra terms present when $n\neq 0$, it does not appear that either of
these constraint equations is a total derivative. \ For example, consider
setting $n=(\frac{\kappa m}{2})^{2}$ so as to simplify the equation as much
as possible. \ In this case eq. (\ref{finalconstraint1}) reduces to%
\begin{equation*}
0=d_{+}\tanh (d_{+}z)\kappa m\sqrt{\gamma }\left( 1+\frac{\sqrt{\gamma }%
\kappa m}{2c_{+}}\tanh (c_{+}(z-L))\right)
\end{equation*}%
This implies either 
\begin{equation*}
d_{+}z=0
\end{equation*}%
or%
\begin{equation*}
\tanh (c_{+}(z-L))=-\frac{2c_{+}}{\kappa m\sqrt{\gamma }}
\end{equation*}%
We can only satisfy the first equation by setting $z=0$. However this
corresponds to both particles at the origin, i.e. a single particle at the
origin with mass $2m$ and no charge; this was previously solved \cite{circle}%
. \ 

Turning to the second equation, since $z$ is independent of time we can
regard this equation as yielding a solution for $\gamma $ as a function of $%
\tau $. \ A problem occurs because there are more constraint equations that
need to be satisfied, and in the uncharged case the other constraint
equations are the equations that determine the time dependence of $\gamma $.

In this case the other constraint equations cannot be satisfied. \ This can
be most easily seen by applying this result to (\ref{finalconstraint-3}). \
Substituting $\tanh (c_{+}(z-L))=-\frac{2c_{+}}{\kappa m\sqrt{\gamma }}$ and 
$n=(\frac{\kappa m}{2})^{2}$ into (\ref{finalconstraint-3}) yields the
following result

\begin{equation*}
\dot{c}_{+}(z-L)+\frac{\gamma \tau \dot{\tau}}{d_{+}^{2}}(\frac{\kappa m%
\sqrt{\gamma }}{2c_{+}})=0
\end{equation*}%
Now $d_{+}^{2}=c_{+}^{2}+\gamma n=c_{+}^{2}+\gamma (\frac{\kappa m}{2}%
)^{2}=c_{+}^{2}(1+(\frac{\kappa m\sqrt{\gamma }}{2c_{+}})^{2})$ , so 
\begin{equation}
\dot{c}_{+}=-\frac{1}{z-L}\frac{\kappa m\sqrt{\gamma }}{2c_{+}}\frac{\gamma
\tau \dot{\tau}}{c_{+}^{2}}\frac{1}{1+(\frac{\kappa m\sqrt{\gamma }}{2c_{+}}%
)^{2}}  \label{cp1}
\end{equation}%
But $\tanh (c_{+}(z-L))=-\frac{2c_{+}}{\kappa m\sqrt{\gamma }}$ gives us a
solution for $c_{+}$ i.e.

\begin{equation*}
c_{+}=\frac{1}{z-L}\tanh ^{-1}\left( \frac{2c_{+}}{\kappa m\sqrt{\gamma }}%
\right) =\frac{1}{z-L}\tanh ^{-1}\left( -\frac{2\sqrt{\tau ^{2}-\Lambda
_{out}/2}}{\kappa m}\right)
\end{equation*}%
which implies that 
\begin{eqnarray}
\dot{c}_{+} &=&-\frac{1}{z-L}\frac{1}{1-(\frac{2c_{+}}{\kappa m\sqrt{\gamma }%
})^{2}}\frac{2}{\kappa m}\frac{\tau \dot{\tau}}{\sqrt{\tau ^{2}-\Lambda
_{out}/2}}  \notag \\
\dot{c}_{+} &=&-\frac{1}{z-L}\frac{1}{(\frac{\kappa m\sqrt{\gamma }}{2c_{+}}%
)^{2}-1}\frac{\kappa m\sqrt{\gamma }}{2c_{+}}\frac{\gamma \tau \dot{\tau}}{%
c_{+}^{2}}  \label{cp2}
\end{eqnarray}%
Comparing eqs. (\ref{cp1}) and (\ref{cp2}) implies that $\frac{2c_{+}}{%
\kappa m\sqrt{\gamma }}=0$ , ie. $c_{+}=0$ which contradicts our initial
assumption that $c_{+}^{2}>0$. \ The same problem occurs when the case $%
c^{2}=c_{-}^{2}<0$ when looking at the case $n=(\frac{\kappa m}{2})^{2.}$

Another simplification occurs if $n=-(\frac{\kappa m}{2})^{2}$, Similar
manipulations yield 
\begin{equation*}
\tanh (d_{+}z)=\frac{2d_{+}}{\kappa m\sqrt{\gamma }}
\end{equation*}%
which is of the same form as the previous result. \ Again this is a problem
because the other constraint equations are not satisfied. \ While these
results don't rule out solutions for all $n\neq 0$ , they illustrate the
problems in finding them. \ \ In fact intuitively one might think that $n=-(%
\frac{\kappa m}{2})^{2}$ and $n=(\frac{\kappa m}{2})^{2}$ provide bounds on $%
n$.\ 

We have been unsuccessful in finding any further ansatze that could solve
the constraint equations. \ Further progress might be attained by attempting
to solve the constraint equations numerically for different values of $n$\
to see if solutions exist for $n\neq 0$. \ However our initial attempts at
solving the equations numerically have uncovered further difficulties: the
problem is that the system appears to be overconstrained when $n\neq 0$. \
When $n=0$\ then (\ref{finalconstraint-2}) and (\ref{finalconstraint-3})
both reduce to the same constraint, but this not the case for arbitrary $n$.

We turn now to consideration of the $n=0$\ case.

\subsection{The $n=0$ solution}

When $n=0$ the previous solutions of $N_{0},\Psi ,N_{1}$ simplify and the
constraint equations derived from (\ref{final-7}) and (\ref{final-8}) are
satisfied. \ The solutions can be written as follows%
\begin{eqnarray}
\Pi &=&\Pi _{0}(t)\exp (\Psi /2),  \label{Pi0} \\
N_{0} &=&\frac{\dot{\tau}\gamma }{\beta c_{+}^{2}}\cosh (c_{+}f(x))-\frac{%
\dot{\tau}\gamma }{c_{+}^{2}}  \label{N0}
\end{eqnarray}%
where%
\begin{equation*}
\beta =\cosh (c_{+}\frac{L}{2})+\frac{4c_{+}}{\kappa m\sqrt{\gamma }}\sinh
(c_{+}\frac{L}{2})
\end{equation*}%
and 
\begin{equation}
f(x)=\LARGE\{ {|x|,|x|<\frac{L}{2}}{L-|x|,|x|>\frac{L}{2}}\LARGE\}
\label{sawtooth}
\end{equation}%
is the saw-tooth function that has magnitude zero halfway between the
particles and magnitude $\frac{L}{2}$ at the locations of the particles. We
also have 
\begin{equation}
N_{1}=\left\{ 
\begin{array}{c}
\gamma \frac{\dot{c}_{+}}{c_{+}}(x+L)-\frac{\gamma ^{2}\dot{\tau}\tau }{%
c_{+}^{3}\beta }\sinh (c_{+}(x+L)),x<-\frac{L}{2} \\ 
\gamma \frac{\dot{c}_{+}}{c_{+}}x-\frac{\gamma ^{2}\dot{\tau}\tau }{%
c_{+}^{3}\beta }\sinh (c_{+}x),-\frac{L}{2}<x<\frac{L}{2} \\ 
\gamma \frac{\dot{c}_{+}}{c_{+}}(x-L)-\frac{\gamma ^{2}\dot{\tau}\tau }{%
c_{+}^{3}\beta }\sinh (c_{+}(x-L)),x>\frac{L}{2}%
\end{array}%
\right\}  \label{N1}
\end{equation}%
\begin{equation}
\Psi =-2\ln (\beta -\cosh (c_{+}f(x)))-2\ln \left( \frac{\kappa \Pi _{0}}{%
c_{+}\sqrt{\beta ^{2}-1}}\right)  \label{Psi}
\end{equation}%
Notice that eq. (\ref{Pi0}) and (\ref{Psi}) together imply%
\begin{equation}
\kappa \Pi =\frac{\pm c_{+}\sqrt{\beta ^{2}-1}}{\beta -\cosh (c_{+}f(x))}
\end{equation}%
and solving (\ref{final-5}) we obtain%
\begin{equation*}
\Pi _{0}=\sqrt{\gamma }\zeta ^{2}\left( \tau \pm \sqrt{\frac{\kappa ^{2}m^{2}%
}{16(\xi ^{2}-1)}+(\tau ^{2}-\Lambda _{e}/2)}\right)
\end{equation*}

Using the solution (\ref{rtgam}) or the metric%
\begin{equation}
\sqrt{\gamma }=\frac{2}{L\sqrt{\tau ^{2}-\Lambda _{e}/2}}\text{arctanh}%
\left( \frac{\xi \sqrt{\frac{\kappa ^{2}m^{2}}{16}+(\xi ^{2}-1)\left( \tau
^{2}-\Lambda _{e}/2\right) }-\frac{\kappa m}{4}}{\frac{\kappa ^{2}m^{2}}{16}%
+\xi ^{2}\left( \tau ^{2}-\Lambda _{e}/2\right) }\sqrt{\left( \tau
^{2}-\Lambda _{e}/2\right) }\right)
\end{equation}%
allows us to rewrite 
\begin{equation}
\beta =\frac{4}{\kappa m}\sqrt{\frac{\kappa ^{2}m^{2}}{16}+(\xi ^{2}-1)(\tau
^{2}-\Lambda _{e}/2)}
\end{equation}

These equations are the solution for when $c^{2}=c_{+}^{2}>0$ but solutions
also exist when $c^{2}=c_{-}^{2}<0$ which can be obtained by substituting $%
c_{+}\rightarrow ic_{-}$ into the above results. The solution for the metric
function then becomes

\begin{equation}
\sqrt{\gamma }=\frac{2}{L\sqrt{\Lambda _{e}/2-\tau ^{2}}}\left[ \arctan
\left( \frac{\xi \sqrt{\frac{\kappa ^{2}m^{2}}{16}+(1-\xi ^{2})\left(
\Lambda _{e}/2-\tau ^{2}\right) }-\frac{\kappa m}{4}}{\frac{\kappa ^{2}m^{2}%
}{16}-\xi ^{2}\left( \Lambda _{e}/2-\tau ^{2}\right) }\sqrt{\left( \Lambda
_{e}/2-\tau ^{2}\right) }\right) +k\pi \right]
\end{equation}%
where $k$ is an integer and $\left| \xi \right| <\sqrt{1+\frac{\left( \kappa
m\right) ^{2}}{8\Lambda _{e}}}$ as a consequence of the periodicity of $%
N_{1}(x,t)$. \ No solutions exist for $c^{2}=0$. \ 

Note that in either case the two particles remain in equilibrium in the
sense that the ratio of the proper lengths between them as computed from
either the rightward or the leftward side of the particle at $z=0$\ is
constant. \ These solutions are very similar to those obtained in the
neutral system \cite{rkern}, \ the key difference being that cosmological
constant has been replaced by $\Lambda _{e}=\Lambda -\kappa E^{2}$ . The
electric field is constant in magnitude but alternates in sign between each
the particles. This result can be easily generalized to $N$ particles.

\section{Charged N-Body Solution}

The only difference between eqs. (\ref{final-1})-(\ref{final-8}) and their
counterparts in the neutral self-gravitating system is the presence of $%
E^{2} $ in equations (\ref{final-2}),(\ref{final-3}) and the term $\int
dxN_{0}\sqrt{\gamma }E\frac{dE}{dz_{a}}$ in equation (\ref{final-7}). \
Using the ansatz $E^{2}=$ constant, it is possible to eliminate $E^{2}$ in
equations (\ref{final-2}),(\ref{final-3}) by introducing $\Lambda
_{e}=\Lambda -\kappa E^{2}$ . \ Furthermore, we have 
\begin{eqnarray*}
\int dxE\frac{dE}{dz_{a}} &=&\int_{z_{a}-\epsilon }^{z_{a}+\epsilon }dxE%
\frac{dE}{dx}=\left. (E^{2})\right| _{z_{a}-\epsilon }^{z_{a}+\epsilon
}-\int_{z_{a}-\epsilon }^{z_{a}+\epsilon }dxE\frac{dE}{dx} \\
\int_{z_{a}-\epsilon }^{z_{a}+\epsilon }dxE\frac{dE}{dx} &=&\frac{1}{2}%
\left. (E^{2})\right| _{z_{a}-\epsilon }^{z_{a}+\epsilon }=0,
\end{eqnarray*}
when $E^{2}$ is constant. \ This implies that $\int dxN_{0}\sqrt{\gamma }E%
\frac{dE}{dz_{a}}=0$, \ since $\int dxN_{0}\sqrt{\gamma }E\frac{dE}{dz_{a}}%
=N_{0}\sqrt{\gamma }\int dxE\frac{dE}{dz_{a}}$ .

So an equilibrium solution for $N$ particles with charge distributed evenly
around the circle exists when $E^{2}$ is constant. \ This yields for $%
c_{+}^{2}>0$

\begin{equation}
\sqrt{\gamma }=\frac{N}{L\sqrt{\tau ^{2}-\Lambda _{e}/2}}\text{arctanh}%
\left( \frac{\xi \sqrt{\frac{\kappa ^{2}M^{2}}{16N^{2}}+(\xi ^{2}-1)(\tau
^{2}-\Lambda _{e}/2)}-\frac{\kappa M}{4N}}{\frac{\kappa ^{2}M^{2}}{16N^{2}}%
+\xi ^{2}(\tau ^{2}-\Lambda _{e}/2)}\sqrt{\tau ^{2}-\Lambda _{e}/2}\right)
\label{gamnbody}
\end{equation}%
where $M=mN$ is the total mass of the system. \ We also have

\begin{eqnarray}
\Psi &=&-2\ln \left( \beta -\cosh (c_{+}f(x))\right) -2\ln (\frac{\kappa \Pi
_{0}\left( t\right) }{c_{+}\sqrt{\beta ^{2}-1}})  \label{psisum} \\
N_{0} &=&\frac{\dot{\tau}\gamma }{c_{+}^{2}\beta }\left( \cosh
(c_{+}f(x))-\beta \right)  \label{lapsesum} \\
N_{1} &=&\gamma \frac{\dot{c}_{+}}{c_{+}}h(x)-\frac{\gamma ^{2}\dot{\tau}%
\tau }{c_{+}^{3}\beta }\sinh (c_{+}h(x))
\end{eqnarray}%
where $f(x)$ is the saw-tooth function that peaks with a value of $L/N$
(i.e. $f(z_{a})=L/N$) at the particle locations and has zero magnitude
half-way between the particles (i.e. $f(\frac{z_{a}+z_{a+1}}{2})=0$ so $f(%
\frac{z_{a}+z_{a+1}}{2}+y)=|y|$ for $|y|<L/N$). The function $h(x)$ is a
function that has a jump discontinuity across the particles, and between two
particles linearly changes from $-L/N$ to $L/N$ (i.e. $h(\frac{z_{a}+z_{a+1}%
}{2}+y)=y$, for $|y|<L/N$). \ Notice that $|h(x)|=f(x)$ and $h^{\prime
}(x)=1 $ between the particles.

The functions $\beta $ and $\Pi _{0\text{ }}$are given by%
\begin{eqnarray}
\beta &=&\frac{4N}{\kappa M}\sqrt{\frac{\kappa ^{2}M^{2}}{16N^{2}}+(\xi
^{2}-1)(\tau ^{2}-\Lambda _{e}/2)}  \label{betanbodysol} \\
\Pi _{0} &=&\sqrt{\gamma }\zeta ^{2}\left( \tau \pm \sqrt{\frac{\kappa
^{2}M^{2}}{16N^{2}(\xi ^{2}-1)}+(\tau ^{2}-\Lambda _{e}/2)}\right)
\label{Pi0nbodysol}
\end{eqnarray}%
where $\zeta $ and $\xi $ are integration constants. \ Then 
\begin{equation}
\kappa \Pi =\frac{\pm c_{+}\sqrt{\beta ^{2}-1}}{\cosh (c_{+}f(x))-\beta }
\label{Pinbody}
\end{equation}

It is possible to write the functions for $\Psi ,N_{0},N_{1}$ in a different
form by using the following identity 
\begin{equation}
\cosh (c_{+}f(x))=\frac{\sinh (\frac{c_{+}L}{N})}{\sinh (c_{+}L)}%
\sum_{a=1}^{N}\cosh (c_{+}(|x-z_{a}|-L))  \label{coshpf}
\end{equation}%
whose proof is given in the appendix. Then for finding $N_{1}$ it is
possible to integrate (\ref{final-4}) to find the following expression \ 
\begin{equation*}
N_{1}=\gamma \frac{\dot{c_{+}}}{c_{+}}x-\frac{\gamma ^{2}\tau \dot{\tau}}{%
c_{+}^{3}\beta }\frac{\sinh (\frac{c_{+}L}{N})}{\sinh (c_{+}L)}%
\sum_{a=1}^{N}\epsilon \left( x-z_{a}\right) \left[ \sinh
(c_{+}(|x-z_{a}|-L))+\sinh (c_{+}L)\right]
\end{equation*}

As with the two particle case there are solutions for $c^{2}=c_{-}^{2}<0$.
These are obtained upon replacing $c_{+}\rightarrow ic_{-}$ in the above
results. \ The solution for the metric function $\gamma $\ now becomes%
\begin{equation}
\sqrt{\gamma }=\frac{N}{L\sqrt{\Lambda _{e}/2-\tau ^{2}}}\left[ \arctan
\left( \frac{\xi \sqrt{\frac{\kappa ^{2}M^{2}}{16N^{2}}+(1-\xi ^{2})(\Lambda
_{e}/2-\tau ^{2})}-\frac{\kappa M}{4N}}{\frac{\kappa ^{2}M^{2}}{16N^{2}}-\xi
^{2}(\Lambda _{e}/2-\tau ^{2})}\sqrt{\Lambda _{e}/2-\tau ^{2}}\right) +k\pi %
\right]  \label{gamnbodyneg}
\end{equation}%
where $k$ is an integer and the integration constant $\left\vert \xi
\right\vert <\sqrt{1+\frac{\left( \kappa M\right) ^{2}}{8\Lambda _{e}N^{2}}}$
as a consequence of the periodicity of $N_{1}(x,t)$.

So an equilibrium solution exists for $N$ charged bodies when when $E^{2}$
is constant. The charges of the particles alternate in sign around the
circle; since the total charge vanishes we have a solution only when $N$ is
an even number.\ 

If the total number of particles is odd then some particles must remain
uncharged. \ The simplest example is the 3-particle case, one particle has
charge $q$, one has charge $-q,$ and one particle has zero charge then it is
possible to choose $E_{0}$ so that $E^{2}=const.$\bigskip

The previous results can be approximated for the case when we take the limit 
$N\rightarrow \infty $ but we keep the total mass of the ring $M$ fixed .

For the case that $c_{+}^{2}>0$ then the solutions can be simplified to:

\begin{eqnarray*}
\sqrt{\gamma } &=&\frac{N}{L\sqrt{\tau ^{2}-\Lambda _{e}/2}}\text{arctanh}%
\left( \frac{\sqrt{(\xi ^{2}-1)}}{\xi }\right) \\
c_{+}^{2} &=&\frac{N^{2}}{L^{2}}\text{arctanh}^{2}\left( \frac{\sqrt{(\xi
^{2}-1)}}{\xi }\right) \\
\beta &=&\frac{4N}{\kappa M}\sqrt{(\xi ^{2}-1)(\tau ^{2}-\Lambda _{e}/2)} \\
\Pi _{0} &=&\sqrt{\gamma }\zeta ^{2}\left( \tau \pm \sqrt{(\tau ^{2}-\Lambda
_{e}/2)}\right)
\end{eqnarray*}

\begin{eqnarray}
\Psi &=&-2\ln \left( \beta -\cosh (c_{+}f(x))\right) -2\ln (\frac{\kappa \Pi
_{0}\left( t\right) }{c_{+}\sqrt{\beta ^{2}-1}}) \\
N_{0} &=&\frac{\dot{\tau}\gamma }{c_{+}^{2}\beta }\left( \cosh
(c_{+}f(x))-\beta \right) \\
N_{1} &=&\gamma \frac{\dot{c}_{+}}{c_{+}}h(x)-\frac{\gamma ^{2}\dot{\tau}%
\tau }{c_{+}^{3}\beta }\sinh (c_{+}h(x))
\end{eqnarray}%
where $f(x)$ is the saw-tooth function that peaks with a value of $L/N$
(i.e. $f(z_{a})=L/N$) at the particle locations and has zero magnitude
half-way between the particles (i.e. $f(\frac{z_{a}-z_{a+1}}{2})=0$ ). \ 

For the case $c_{-}^{2}<0$ the solution for the metric function $\gamma $\
becomes

\begin{eqnarray*}
\sqrt{\gamma } &=&\frac{N}{L\sqrt{\Lambda _{e}/2-\tau ^{2}}}\left[ \arctan
\left( \frac{\sqrt{(1-\xi ^{2})}}{\xi }\right) +k\pi \right] \\
c_{-}^{2} &=&-\frac{N^{2}}{L^{2}}\left[ \arctan \left( \frac{\sqrt{(1-\xi
^{2})}}{\xi }\right) +k\pi \right] ^{2}
\end{eqnarray*}

where $k$ is an integer and the integration constant $\left\vert \xi
\right\vert <1$

\section{Summary of N-body Solutions}

The 2-body equilibrium solution we found is simply a special case of the
N-body equilibrium solution that occurs when $E^{2}=$constant. \ For
convenience here is a table summarizing the results for $c_{+}^{2}>0$:

\begin{center}
\begin{tabular}{|l|l|}
\hline
variable & solution \\ \hline
$\Pi $ & $\Pi _{0}(t)\exp (\Psi /2)$ \\ \hline
$\Psi $ & $-2\ln \left( \beta -\cosh (c_{+}f(x))\right) -2\ln (\frac{\kappa
\Pi _{0}}{c_{+}\sqrt{\beta ^{2}-1}})$ \\ \hline
$N_{0}$ & $\frac{\dot{\tau}\gamma }{c_{+}^{2}\beta }\left( \cosh
(c_{+}f(x))-\beta \right) $ \\ \hline
$N_{1}$ & $\gamma \frac{\dot{c}_{+}}{c_{+}}h(x)-\frac{\gamma ^{2}\dot{\tau}%
\tau }{c_{+}^{3}\beta }\sinh (c_{+}h(x))$ \\ \hline
$\Pi _{0}$ & $\sqrt{\gamma }\zeta ^{2}\left( \tau \pm \sqrt{\frac{\kappa
^{2}M^{2}}{16N^{2}(\xi ^{2}-1)}+(\tau ^{2}-\Lambda _{e}/2)}\right) $ \\ 
\hline
$\beta $ & $\frac{4N}{\kappa M}\sqrt{\frac{\kappa ^{2}M^{2}}{16N^{2}}+(\xi
^{2}-1)(\tau ^{2}-\Lambda _{e}/2)}$ \\ \hline
$c_{+}^{2\text{ }}$ & $\gamma (\tau ^{2}-\Lambda _{e}/2)\text{ }$ \\ \hline
$\sqrt{\gamma }$ & $\frac{N}{L\sqrt{\tau ^{2}-\Lambda _{e}/2}}$arctanh$%
\left( \frac{\xi \sqrt{\frac{\kappa ^{2}M^{2}}{16N^{2}}+(\xi ^{2}-1)(\tau
^{2}-\Lambda _{e}/2)}-\frac{\kappa M}{4N}}{\frac{\kappa ^{2}M^{2}}{16N^{2}}%
+\xi ^{2}(\tau ^{2}-\Lambda _{e}/2)}\sqrt{\tau ^{2}-\Lambda _{e}/2}\right) $
\\ \hline
$f(x)$ & $f(\frac{z_{a}+z_{a+1}}{2}+y)=|y|$ for $|y|<L/N$ \\ \hline
$h(x)$ & $h(\frac{z_{a}+z_{a+1}}{2}+y)=y$, for $|y|<L/N$ \\ \hline
$\Lambda _{e}$ & $\Lambda -\kappa E^{2}$ \\ \hline
\end{tabular}
\end{center}

For $c^{2}=c_{-}^{2}<0$ replace $c_{+}\rightarrow ic_{-}$ and the metric is
now

\begin{center}
\begin{tabular}{|l|l|}
\hline
variable & solution \\ \hline
$N_{0}$ & $-\frac{\dot{\tau}\gamma }{c_{-}^{2}\beta }\left( \cos
(c_{-}f(x))-\beta \right) $ \\ \hline
$N_{1}$ & $\gamma \frac{\dot{c}_{+}}{c_{+}}h(x)+\frac{\gamma ^{2}\dot{\tau}%
\tau }{c_{-}^{3}\beta }\sin (c_{-}h(x))$ \\ \hline
$\sqrt{\gamma }$ & $\frac{N}{L\sqrt{\Lambda _{e}/2-\tau ^{2}}}\left[ \arctan
\left( \frac{\xi \sqrt{\frac{\kappa ^{2}M^{2}}{16N^{2}}+(1-\xi ^{2})(\Lambda
_{e}/2-\tau ^{2})}-\frac{\kappa M}{4N}}{\frac{\kappa ^{2}M^{2}}{16N^{2}}-\xi
^{2}(\Lambda _{e}/2-\tau ^{2})}\sqrt{\Lambda _{e}/2-\tau ^{2}}\right) +k\pi %
\right] $ \\ \hline
$\xi $ & $\left| \xi \right| <\sqrt{1+\frac{\left( \kappa M\right) ^{2}}{%
8\Lambda _{e}N^{2}}}$ \\ \hline
\end{tabular}
\end{center}

\noindent where $k$ is an integer.

\section{Discussion}

We have obtained an exact equilibrium solution for $N$ charged particles of
equal mass in circular gravity. \ Our notion of equilibrium has been taken
to be that of a static equilibrium: namely that the ratio of the proper
lengths between any two pairs of particles is constant. On any $t=$constant
hypersurface this implies that the values of the $z_{a}$\ are constant. \ 

Our solution is valid only when the electric field is of constant magnitude.
\ The electric field will be constant when the number of charged particles
is even and these particles must have the same magnitude charge alternating
in sign between positive and negative around the circle. \ Although
restrictive, there is still a lot of freedom in choosing the arrangement of
the charged particles situated amongst any number of neutral particles that
are present between them. \ For example it is possible to have a system of $%
N $\ particles with only two charged -- then there would be $N-1$\ ways to
arrange the charged particles within the system. \textbf{\ }

Apart from these distinctions, our solution is otherwise the same as that
obtained in the electrically neutral case \cite{rkern}. \ Our solutions
almost certainly describe an unstable equilibrium, since the masses are all
equal and the particles are evenly separated. \ In fact an $N$-body stable
equilibrium does exist within the context of our restrictions, but
corresponds to all particles on top of each other, which is simply the one
body solution.

We were unable to obtain the most general analytic solutions\textbf{\ }to
the full set of equations. An inspection of the constraint equations
suggests that they can only be simultaneously satisfied under very stringent
circumstances. Indeed, the situation with two bodies of unequal mass is
quite perplexing. One might expect that a solution of unstable equilibrium
would exist for two bodies located equidistant from one another on the
circle. \ However our rather general ansatz (\ref{ABansatz}) yields a
consistent solution to all equations except for that which governs the
time-evolution of $\Pi $ (eq. (\ref{fin13})); we can find solutions to this
last equation only when the masses are equal. \ Similar stringent
constraints hold in the charged cases, where again we were able to find
solutions only for equal masses.

It would appear that the most promising way to make further progress would
be to solve the equations of $(1+1)$ circular gravity numerically. \ Lineal
self-gravitating systems have been studied numerically and interesting
results have been obtained in both the 2-body \cite{2bdchglo} and 3-body %
\cite{fiona}\ cases. Currently there have been no numerical models of $(1+1)$%
\ circular gravity. \ It would be interesting to numerically model two
particles on a circle and to see how the space-time itself behaves
differently when compared to the behavior of two-body equilibrium solution.
\ It would also be interesting to numerically model the system of particles
when the masses are unequal or when charge has been introduced or both. \ 

{\Large Acknowledgements}

This work was supported by the Natural Sciences and Engineering Research
Council of Canada.\ 

\section{Appendix}

\subsection{Derivation of Equation (98)}

In examining equation (\ref{final-7}) notice that $N_{0}^{\prime }(z_{a})$
needs to be defined since $N_{0}^{\prime }(x)$ is discontinuous across the
particles. This has been seen before in equation (\ref{N0-con2}) 
So $N_{0}^{\prime }(z_{a})$ is defined as 
\begin{equation}
N_{0}^{\prime }(z_{a})\equiv \lim_{\epsilon \rightarrow 0}\frac{%
N_{0}^{\prime }(z_{a}-\epsilon )+N_{0}^{\prime }(z_{a}+\epsilon )}{2}
\end{equation}%
Now examining the $\int dxN_{0}\sqrt{\gamma }E\frac{dE}{dz_{a}}$ term of
equation it is useful to rewrite this term as%
\begin{eqnarray*}
\int dxN_{0}\sqrt{\gamma }E\frac{dE}{dz_{a}} &=&N_{0}\sqrt{\gamma }%
\int_{z_{a}-\epsilon }^{z_{a}+\epsilon }dx\left( \frac{1}{2}%
\sum_{b}e_{b}\epsilon (x-z_{b})+E_{0}\right) \left( \sum_{a}e_{a}\delta
(x-z_{a})\right) \\
&=&N_{0}(z_{a})\sqrt{\gamma }e_{a}\left( \frac{1}{2}\sum_{b}e_{b}\epsilon
(z_{a}-z_{b})+E_{0}\right)
\end{eqnarray*}%
Returning to the two particle example, this implies that at $x=-z$

\begin{eqnarray*}
\int dxN_{0}\sqrt{\gamma }E\frac{dE}{d(-z)} &=&N_{0}(-z)\sqrt{\gamma }\frac{1%
}{2}q\left( q+2E_{0}\right) \\
&=&N_{0}(-z)\sqrt{\gamma }\frac{n}{\kappa }
\end{eqnarray*}%
and at $x=+z$

\begin{eqnarray*}
\int dxN_{0}\sqrt{\gamma }E\frac{dE}{dz} &=&-N_{0}(z)\sqrt{\gamma }\frac{1}{2%
}q\left( q+2E_{0}\right) \\
&=&-N_{0}(z)\sqrt{\gamma }\frac{n}{\kappa }
\end{eqnarray*}

\bigskip Note that because of the symmetry properties of $N_{0}$ we have $%
N_{0}^{\prime }(-z)=-N_{0}^{\prime }(z)$ and $\int dxN_{0}\sqrt{\gamma }E%
\frac{dE}{d(-z)}=-\int dxN_{0}\sqrt{\gamma }E\frac{dE}{dz}$. \ So (\ref%
{final-7}) is the same equation at $z$ and $-z$ and therefore it is only a
single constraint equation. \ Explicitly $N_{0}^{\prime }(z)$ is given by%
\begin{equation*}
N_{0}^{\prime }(z)=\frac{1}{2}\left\{ A\left[ d_{+}\sinh (d_{+}z)+c_{+}\cosh
(d_{+}z)\tanh (c_{+}(z-L))\right] +\frac{\dot{\tau}\gamma ^{2}n}{%
c_{+}d_{+}^{2}}\tan (c_{+}(z-L))\right\}
\end{equation*}%
\bigskip where%
\begin{eqnarray*}
A &=&\left( \frac{\dot{\tau}\gamma }{Bd_{+}^{2}}\right) \left( \frac{\kappa m%
\sqrt{\gamma }}{2}+\frac{\gamma n}{c_{+}}\tanh (c_{+}(z-L))\right) \\
\frac{\dot{\tau}\gamma }{d_{+}^{2}} &=&\frac{AB}{\left( \frac{\kappa m\sqrt{%
\gamma }}{2}+\frac{\gamma n}{c_{+}}\tanh (c_{+}(z-L))\right) }
\end{eqnarray*}%
and so%
\begin{eqnarray}
N_{0}^{\prime }(z) &=&\frac{A\cosh (d_{+}z)}{2\left( \frac{\kappa m\sqrt{%
\gamma }}{2}+\frac{\gamma n}{c_{+}}\tanh (c_{+}(z-L))\right) }\left\{ \frac{%
\kappa m\sqrt{\gamma }}{2}d_{+}\tanh (d_{+}z)\right.  \notag \\
&&+\frac{\kappa m\sqrt{\gamma }}{2}c_{+}\tanh (c_{+}(z-L))\left[ 1+\frac{%
\gamma n}{c_{+}^{2}}\right]  \notag \\
&&\left. +2\frac{\gamma n}{c_{+}}d_{+}\tanh (d_{+}z)\tanh
(c_{+}(z-L))\right\}  \label{N0prime1}
\end{eqnarray}%
\bigskip

The constraint equation (\ref{final-7}) implies that%
\begin{equation*}
N_{0}^{\prime }(z)=\frac{\sqrt{\gamma }n}{m\kappa }\left( A\cosh (d_{+}z)-%
\frac{\dot{\tau}\gamma }{d_{+}^{2}}\right)
\end{equation*}%
or alternatively%
\begin{eqnarray}
N_{0}^{\prime }(z) &=&\frac{A\cosh (d_{+}z)}{2\left( \frac{\kappa m\sqrt{%
\gamma }}{2}+\frac{\gamma n}{c_{+}}\tanh (c_{+}(z-L))\right) }\left\{ \frac{%
2\gamma n}{\sqrt{\gamma }m\kappa }c_{+}\tanh (c_{+}(z-L))\left[ \frac{\gamma
n}{c_{+}^{2}}+1\right] \right.  \notag \\
&&\left. -\frac{2\gamma n}{\sqrt{\gamma }m\kappa }d_{+}\tanh (d_{+}z)\right\}
\label{N0prime2}
\end{eqnarray}

Since the constraint (\ref{N0prime2}) must be equivalent to the definition
of $N_{0}^{\prime }(z)$ in eq. (\ref{N0prime1}) we have%
\begin{eqnarray}
0 &=&d_{+}\tanh (d_{+}z)\frac{2\sqrt{\gamma }}{m\kappa }\left[ n+(\frac{%
\kappa m}{2})^{2}\right] +2\frac{\gamma n}{c_{+}}d_{+}\tanh (d_{+}z)\tanh
(c_{+}(z-L))  \notag \\
&&+c_{+}\tanh (c_{+}(z-L))\left[ \frac{2\gamma ^{\frac{3}{2}}n}{m\kappa
c_{+}^{2}}\left( (\frac{\kappa m}{2})^{2}-n\right) +\frac{2\sqrt{\gamma }}{%
m\kappa }\left( (\frac{\kappa m}{2})^{2}-n\right) \right]
\end{eqnarray}%
which is the constraint equation that results from (\ref{final-7}).

\subsection{Proof of Identity (121)}

Now returning to the $N$-Body equilibrium solution when $E^{2}=const$ an
interesting identity that was mentioned without proof was eq. (\ref{coshpf}%
). \ This identity is useful because it allows us to write the functions $%
\Psi ,N_{0},N_{1}$ in two different forms: either as a single function or as
a sum of $N$ functions. \ Without this identity it would not be apparent
that both forms of the equations are in fact equivalent. \ \ 

We now verify eq. (\ref{coshpf}). We begin by noting that both $\cosh
(c_{+}f(x))$ and $\frac{\sinh (\frac{c_{+}L}{N})}{\sinh (c_{+}L)}%
\sum_{a=1}^{N}\cosh (c_{+}(|x-z_{a}|-L))$ are periodic between the
particles. \ This is equivalent to saying that both functions have a period
of $\frac{2L}{N}$. \ For $\cosh (c_{+}f(x))$ this can be seen because the
sawtooth function $f(x)$ peaks at the particles, yielding $f(x+\frac{2L}{N}%
)=f(x)$. \ In fact there are two values for each $|x-z_{a}|$ because this
system is located on a circle. \ Thus there are two possible directions in
which one can measure the distance between $x$ and $z_{a}$. If $|x-z_{a}|$
is the distance between them one way around the circle then $2L-|x-z_{a}|$
is the distance as measured via the other way. \ Note that regardless of how 
$|x-z_{a}|$ is measured the\ functions $\cosh (c_{+}(|x-z_{a}|-L))$ are well
defined. \ This is because 
\begin{eqnarray*}
\cosh (c_{+}(|x-z_{a}|-L)) &=&\cosh (c_{+}([2L-|x-z_{a}|]-L)) \\
&=&\cosh (c_{+}(L-|x-z_{a}|))
\end{eqnarray*}

Since the particles $z_{a}$ are evenly spaced with spacing $\frac{2L}{N}$ it
follows that $\sum_{a=1}^{N}\cosh (c_{+}(|x-z_{a}|-L))$ must be periodic
with period $\frac{2L}{N}$. \ This is because a shift by $\frac{2L}{N}$ can
be accounted for by relabeling the particles i.e.

\begin{eqnarray*}
\sum_{a=1}^{N}\cosh (c_{+}(|x+\frac{2L}{N}-z_{a}|-L)) &=&\sum_{a=1}^{N}\cosh
(c_{+}(|x-(z_{a}-\frac{2L}{N})|-L)) \\
&=&\sum_{b=1}^{N}\cosh (c_{+}(|x-z_{b}|-L))
\end{eqnarray*}

Therefore because both $\cosh (c_{+}f(x))$ and $\frac{\sinh (\frac{c_{+}L}{N}%
)}{\sinh (c_{+}L)}\sum_{a=1}^{N}\cosh (c_{+}(|x-z_{a}|-L))$ are periodic
between the particles it is only necessary to show that these functions are
equivalent between two particles in order to prove that they are equivalent
on the entire circle. \ To prove the equivalence it is necessary to separate
the cases when $N$ is even and when $N$ is odd and prove these cases
independently.

\begin{enumerate}
\item Even $N$
\end{enumerate}

For the case when $N$ is even the origin is located halfway between two
particles, and so the particles are located at $\frac{2k-1}{N}L$ for $k=-%
\frac{N}{2}+1,-\frac{N}{2}+2,...,$ $\frac{N}{2}$. \ We will show equivalence
of $\cosh (c_{+}f(x))$ and $\frac{\sinh (\frac{c_{+}L}{N})}{\sinh (c_{+}L)}%
\sum_{a=1}^{N}\cosh (c_{+}(|x-z_{a}|-L))$ when $\frac{-L}{N}\leq x\leq \frac{%
L}{N}$. \ Recall that the sawtooth function $f(x)$ is defined to have value $%
\frac{L}{N}$ at the particles (i.e. $f(\frac{2k-1}{N}L)=\frac{L}{N}$) and
the value $0$ halfway between the particles (i.e. $f(\frac{2k}{N}L)=0$). \
So this means 
\begin{equation*}
f(x)=|x|\text{ for }\frac{-L}{N}\leq x\leq \frac{L}{N}
\end{equation*}

Now%
\begin{eqnarray*}
\cosh (c_{+}|x|) &=&\cosh (c_{+}x) \\
\therefore \cosh (c_{+}f(x)) &=&\cosh (c_{+}x)\text{ for }\frac{-L}{N}\leq
x\leq \frac{L}{N}
\end{eqnarray*}%
and so it is only required to show that $\sum_{a=1}^{N}\cosh
(c_{+}(|x-z_{a}|-L))=\frac{\sinh (c_{+}L)}{\sinh (\frac{c_{+}L}{N})}\cosh
(c_{+}x)$. \ This is straightforward

\begin{eqnarray*}
&&\sum_{a=1}^{N}\cosh (c_{+}(|x-z_{a}|-L)) \\
&=&\sum_{k=1}^{\frac{N}{2}}\cosh (c_{+}(|x-\frac{2k-1}{N}L|-L))+\sum_{k=-%
\frac{N}{2}+1}^{0}\cosh (c_{+}(|x-\frac{2k-1}{N}L|-L)) \\
&=&\sum_{k=1}^{\frac{N}{2}}\cosh (c_{+}(\frac{2k-1}{N}L-x-L))+\sum_{k=-\frac{%
N}{2}+1}^{0}\cosh (c_{+}(x-\frac{2k-1}{N}L-L)) \\
&=&\sum_{k=1}^{\frac{N}{2}}\cosh (c_{+}(\frac{2k-(N+1)}{N}L-x))+\cosh (c_{+}(%
\frac{2k-(N+1)}{N}L+x))
\end{eqnarray*}

\begin{equation}
\therefore \sum_{a=1}^{N}\cosh (c_{+}(|x-z_{a}|-L))=\sum_{k=1}^{\frac{N}{2}%
}2\cosh (c_{+}x)\cosh (\frac{c_{+}L}{N}(2k-(N+1)))  \label{proofcosh1}
\end{equation}%
and note 
\begin{eqnarray*}
&&\sum_{k=1}^{\frac{N}{2}}\cosh (\frac{c_{+}L}{N}(2k-(N+1))) \\
&=&\frac{1}{2}\sum_{k=1}^{\frac{N}{2}}\left[ e^{\frac{c_{+}L}{N}%
(2k-(N+1))}+e^{-\frac{c_{+}L}{N}(2k-(N+1))}\right] \\
&=&\sum_{k=1}^{\frac{N}{2}}\frac{\sinh (\frac{c_{+}L}{N}(2k-N))-\sinh (\frac{%
c_{+}L}{N}(2k-(N+2)))}{e^{\frac{c_{+}L}{N}}-e^{-\frac{c_{+}L}{N}}} \\
&=&\frac{1}{2\sinh (\frac{c_{+}L}{N})}\left[ \sum_{k=1}^{\frac{N}{2}}\sinh (%
\frac{c_{+}L}{N}(2k-N))-\sum_{k=1}^{\frac{N}{2}}\sinh (\frac{c_{+}L}{N}%
(2(k-1)-N))\right] \\
&=&\frac{\sinh (c_{+}L)}{2\sinh (\frac{c_{+}L}{N})}
\end{eqnarray*}%
So we have%
\begin{equation}
\therefore \sum_{k=1}^{\frac{N}{2}}\cosh (\frac{c_{+}L}{N}(2k-(N+1)))=\frac{%
\sinh (c_{+}L)}{2\sinh (\frac{c_{+}L}{N})}  \label{proofidentity}
\end{equation}%
and by substituting (\ref{proofidentity}) into (\ref{proofcosh1}) we obtain

\begin{equation*}
\sum_{a=1}^{N}\cosh (c_{+}(|x-z_{a}|-L))=\frac{\sinh (c_{+}L)}{\sinh (\frac{%
c_{+}L}{N})}\cosh (c_{+}x)
\end{equation*}

Hence for an even number of particles 
\begin{equation*}
\cosh (c_{+}f(x))=\frac{\sinh (\frac{c_{+}L}{N})}{\sinh (c_{+}L)}%
\sum_{a=1}^{N}\cosh (c_{+}(|x-z_{a}|-L))
\end{equation*}

\begin{enumerate}
\item[2.] N is odd
\end{enumerate}

When the number of particles is odd the origin is located to be at the
location of one of the particles, so the particles are located at $\frac{2k}{%
N}L$ for $k=-\frac{N-1}{2},-\frac{N-1}{2}+1,...,$ $\frac{N-1}{2}$. \ This
time we will show the equivalence of $\cosh (c_{+}f(x))$ and $\frac{\sinh (%
\frac{c_{+}L}{N})}{\sinh (c_{+}L)}\sum_{a=1}^{N}\cosh (c_{+}(|x-z_{a}|-L))$
in the region $0\leq x\leq \frac{2L}{N}$.

So%
\begin{equation*}
f(0)=f(\frac{2L}{N})=\frac{L}{N}
\end{equation*}

\begin{equation*}
\therefore f(x)=\left\{ 
\begin{array}{c}
\frac{L}{N}-x\text{ for }0\leq x\leq \frac{L}{N} \\ 
x-\frac{L}{N}\text{ for }\frac{L}{N}\leq x\leq \frac{2L}{N}%
\end{array}%
\right.
\end{equation*}%
Note%
\begin{eqnarray*}
&&\cosh (c_{+}(\frac{L}{N}-x))=\cosh (c_{+}(x-\frac{L}{N})) \\
&\therefore &\cosh (f(x))=\cosh (c_{+}(x-\frac{L}{N}))\text{ for }0\leq
x\leq \frac{2L}{N}
\end{eqnarray*}%
By a similar calculation to the even case we can show that $%
\sum_{a=1}^{N}\cosh (c_{+}(|x-z_{a}|-L))=\frac{\sinh (c_{+}L)}{\sinh (\frac{%
c_{+}L}{N})}\cosh (c_{+}(x-\frac{L}{N}))$.

Hence when $N$ is odd 
\begin{equation*}
\cosh (c_{+}f(x))=\frac{\sinh (\frac{c_{+}L}{N})}{\sinh (c_{+}L)}%
\sum_{a=1}^{N}\cosh (c_{+}(|x-z_{a}|-L))
\end{equation*}

\end{document}